\documentclass[twocolumn]{aa} 

\usepackage{graphicx}
\usepackage[varg]{txfonts}
\usepackage{color}
\usepackage{xcolor}
\usepackage[colorlinks=black, linkcolor=black, citecolor=black]{hyperref} 

\begin{document} 
 

\def\KMS{km s$^{-1}$}
\def\VLSR{V$_{lsr}$}
\def\TA{T$_{A}^{*}$}
\def\TMB{T$_{mb}$}
\def\TREC{$T_{REC}$}
\def\DEG{$^{\circ}$}
\def\HII{\mbox{H \footnotesize{II}\normalsize}}


\def\CO10{\mbox{CO(1-0)}}
\def\CO21{\mbox{CO(2-1)}}
\def\CO32{\mbox{CO(3-2)}}
\def\COLINEAI{\mbox{CO(4-3)}}
\def\CO65{\mbox{CO(6-5)}}
\def\CO76{\mbox{CO(7-6)}}

\def\CS21{\mbox{CS(2-1)}}
\def\HCN10{\mbox{HCN(1-0)}}
\def\METHANOL{\mbox{CH$_{3}$OH}}
\def\WATER{\mbox{H$_{2}$O}}
\def\AMONIA{\mbox{NH$_{3}$(1,1)}}

\def\COLINEAIl{\mbox{CO(J$=$4 - 3)}}
\def\COLINEAIIl{\mbox{CO(J$=$7 - 6)}}
\def\DIAZENYLIUM{\mbox{N$_{2}$H$^{+}$(1-0)}}

\def\CILINEAI{\mbox{[CI](1-0)}}
\def\CILINEAII{\mbox{[CI](2-1)}}
\def\NII{\mbox{[NII]}}
\def\CII{\mbox{[CII]}}

\def\CILINEAIl{\mbox{[C\footnotesize{I}\normalsize] ${^{3}}$P$_{1}$ - ${^{3}}$P$_{0}$}}
\def\CILINEAIIl{\mbox{[C\footnotesize{I}\normalsize] ${^{3}}$P$_{2}$ - ${^{3}}$P$_{1}$}}
\def\NIIl{\mbox{[N\footnotesize{II}\normalsize] ${^{3}}$P$_{1}$ - ${^{3}}$P$_{0}$}}
\def\CIIl{\mbox{[C\footnotesize{II}\normalsize] ${^{2}}$P$_{3/2}$ - ${^{2}}$P$_{1/2}$}}

\title{ Warm ISM in the Sgr A Complex. I.}
\subtitle{Mid-J CO, atomic carbon, ionized atomic carbon, \\ 
and ionized nitrogen sub-mm/FIR line observations with \\
the Herschel-HIFI and NANTEN2/SMART telescopes}

\author{P. Garc\'ia \inst{1}
\and
R. Simon\inst{1}
\and 
J. Stutzki\inst{1}
\and 
R. G\"usten\inst{2}
\and 
M. A. Requena-Torres\inst{2}
\and 
R. Higgins\inst{1}
}
\institute{I. Physikalisches Institut der Universit\"at zu K\"oln, D-50937 Cologne, Germany.\\
\email{pablo@ph1.uni-koeln.de}
\and
Max-Planck-Institut f\"ur Radioastronomie, Auf dem H\"ugel 69, 53121 Bonn, Germany.\\
}

\date{Received xxxxxx xx, XXXX; Accepted xxxxxx xx, XXXX}

\abstract
{}
{We investigate the spatial and spectral distribution of the LSR (local standard of rest) velocity resolved submillimetre emission from the warm (25 - 90 K) gas in the Sgr A Complex, 
located in the Galactic Centre.}
{We present large-scale submillimetre heterodyne observations towards the Sgr A Complex covering $\sim$ 300 arcmin$^{2}$. These data were obtained in the frame 
of the Herschel EXtraGALactic Guaranteed Time Key Program (HEXGAL) with the Herschel-HIFI satellite and are complemented with submillimetre 
observations obtained with the NANTEN2/SMART telescope as part of the NANTEN2/SMART Central Nuclear Zone Survey. The observed species are \COLINEAIl\textrm{ }at 
461.0 GHz observed with the NANTEN2/SMART telescope, and \CILINEAIl\textrm{ }at 492.2 GHz, \CILINEAIIl\textrm{ }at 809.3 GHz, \NIIl\textrm{ }at 1461.1 GHz, and 
\CIIl\textrm{ }at 1900.5 GHz observed with the Herschel-HIFI satellite. The observations are presented in a 1 \KMS\textrm{ }spectral resolution and a spatial 
resolution ranging from 46 arcsec to 28 arcsec. The spectral coverage of the three lower frequency lines is $\pm$ 200 \KMS, while in the 
two high frequency lines, the upper LSR velocity limit is $+$94 \KMS\textrm{ }and $+$145 \KMS\textrm{ }for the \NII\textrm{ }and \CII\textrm{ }lines, respectively.}
{The spatial distribution of the emission in all lines is very widespread. The bulk of the carbon monoxide emission is found towards Galactic latitudes 
below the Galactic plane, and all the known molecular clouds are identified. Both neutral atomic carbon lines have their brightest emission associated with 
the $+$50 \KMS\textrm{ }cloud. Their spatial distribution at this LSR velocity describes a crescent-shape structure, which is probably the result of interaction 
with the energetic event (one or several supernovae explosions) that gave origin to the non-thermal Sgr A-East source. The \CII\textrm{ }and 
\NII\textrm{ }emissions have most of their flux associated with the thermal Arched-Filaments and the H Region and bright spots in \CII\textrm{ }emission towards the 
Central Nuclear Disk (CND) are detected. Warm Gas at very high ($\mid$\VLSR$\mid$ $>$ 100 \KMS) LSR velocities is also detected towards the line of sight to the Sgr A 
Complex, and it is most probably located outside the region, in the X$_{1}$ orbits.}
{}


   \maketitle


\section{Introduction}\label{Intro}

The Galactic Centre (GC), located at 8.34 $\pm$ 0.16 kpc \citep{reid2014} from the Sun in the Sagittarius (Sgr) constellation, is one of the most 
remarkable places in the Milky Way; it  harbours an overwhelming variety of astronomical sources under extraordinary physical conditions. 
The region between l $=$ $-$1\DEG.5 to 3\DEG.5 and b $= \pm$ 0\DEG.75, also known as the Nuclear Bulge 
\citep{mezger1996}, harbours five  giant molecular clouds (GMCs): Sgr A, Sgr B, Sgr C, Sgr D, and Bania's Clump 2 \citep{oka1998,bania1980,stark1986}. 
The densest GMCs within this region are located in the so-called Central Molecular Zone (CMZ) extending from l $=$ $-$1\DEG.0 to $+$1\DEG.5\textrm{ }and 
containing all GMCs in the Sgr A region.\\

The CMZ has been observed in almost the entire electro-magnetic spectrum, ranging from centimetre wavelengths \citep{yusef1984,law2008}, 
millimetre wavelengths in a wide variety of molecules \citep{nagayama2007,ott2014,jones2013,jones2012,bally1987,oka1998,sawada2001,oka2012,oka2007}, 
atomic, molecular, and dust observations in the submillimetre (sub-mm) range \citep{martin2004,tanaka2011,pierce2000,schuller2009,bally2010},
far- and mid-infrared observations accessible only from space \citep{molinari2011,stolovy2006}, optical imaging \citep{figer2004}, and in the 
high energy range from a few tenths of keV up to TeV \citep{watson1981,sidoli2001,hunter1997,aharonian2006} among others, and at a wide
range of spectral and spatial resolutions. This large amount of astronomical data has revealed the complex kinematics and extreme physical 
conditions in the Nuclear Bulge (see \citet{mezger1996} for a detailed review).\\ 

Among all GMCs in the CMZ, the Sgr A Complex is the one containing the dynamical centre of the Milky Way. The bolometric luminosity of the central parsec in the 
GC ($\sim$ 10$^{8}$ L$_{\odot}$) accounts for $\sim$ 0.3\% of the bolometric luminosity of the Galaxy. Given the low luminosity of the GC, it is thought that the 
Milky Way would resemble a weak Seyfert Galaxy as seen from a distance of 700 kpc, with an angular resolution of 0.05'' \citep{mezger1996}. Understanding 
the physical conditions and dynamics of the Sgr A Complex is crucial to understanding unresolved CMZs in other Galaxies.\\ 

The Sgr A region, extending $\sim$ 100 pc around Sgr A$^{\star}$, contains a rich diversity of prominent astronomical sources confined in a small volume,
whose emission can be detected across the entire spectral domain. Some of these sources are the non-thermal shell Sgr A-East and the \HII\textrm{ }region 
Sgr A-West \citep{yusef1987a}; the Central Nuclear Disk (CND) \citep{requena2012}; the Radio Arc and several non-thermal filaments (NTFs) 
running perpendicular to the Galactic plane and possibly associated with the poloidal structure of the magnet-field in this region 
\citep{yusef1987b,lang1999b,lang2010}; the Thermal Arched  Filaments \citep{serabyn1987,lang2002}; the H Region containing several \HII\textrm{ }regions 
\citep{lang2001,yusef1987a,zhao1993}; the Sickle and Pistol star \citep{yusef1987b,timmermann1996,lang1997}; the three massive star clusters 
Quintuplet, Arches, and the Nuclear Cluster \citep{cotera1996,lang1999a,chatzopoulos2014}; several molecular clouds (MCs) such as the $-$30 \KMS, $+$20 \KMS, and $+$50 
\KMS\textrm{ }clouds \citep{guesten1981,serabyn1987,zhao1993}; water and methanol masers \citep{caswell2010,caswell2011}; 
X-ray point-like sources such as 1E-1743.1-2843B \citep{porquet2003}; and high velocity compact clumps (HVCCs) \citep{oka2008}. Along the same line of sight (l.o.s.) 
but outside this region are found gas at high LSR (local standard of rest) velocities associated with X$_{1}$ orbits and absorption features at LSR velocities $-$55 \KMS, 
$-$30 \KMS, and $-$5 \KMS\textrm{ }associated with the loci of the 3 kpc, 4.5 kpc, and local arm \citep[Luna et al., in preparation]{oka1998,jones2012}.\\  

In the sub-mm/far-infrared (FIR) range (100 GHz - 10 THz), great effort has been made to understand the 
physics behind the Sgr A Complex and the numerous astronomical sources within it. On the one hand, previous heterodyne observations of this region, 
ranging from a few targeted positions with moderate spectral resolution to large-scale observations with moderate spatial resolution, have
largely contributed to understanding the kinematics and excitation mechanisms of the warm gas in the GC \citep{genzel1990,poglitsch1991,mizutani1994,martin2004}. 
More recently, small-scale observations  with high spatial and spectral resolutions have become available for the shorter wavelengths in this spectral 
domain with the Stratospheric Observatory for Infrared Astronomy (SOFIA) as shown in the observations of the CND by \citet{requena2012}. 
Additionally, several high resolution continuum observations covering the whole CMZ have been crucial to estimating volume densities and temperatures and
to characterizing the spatial morphology of the warm gas in this region \citep{mezger1989,pierce2000,bally2010,molinari2011}.\\

Observation with ground-based telescopes in this frequency domain are particularly difficult given the poor atmospheric transmission. 
In the present work we present observations of the entire Sgr A Complex at high angular ($\leq$ 46'') and spectral (1 \KMS) 
resolution in most of the main cooling lines (except for the two [OI] transitions at 146 $\mu m$ and 63 $\mu m$) in the sub-mm/FIR regime of the 
ISM (\COLINEAI, \CILINEAI, \CILINEAII, and \CII) and \HII\textrm{ }regions (\NII) obtained with the Herschel-HIFI satellite and the NANTEN2/SMART telescope. 
In Section \ref{data_reduction}, the data acquisition and data reduction process are described. In Section \ref{sources} the main sources detected 
in the observations are discussed. In Section \ref{summary} we summarize the work described in the previous sections and outline the future
work in terms of the excitation analysis (PDRs vs XDRs, LVG analysis, etc.) of the gas associated with the variety of sources in our observations 
which are under very different physical conditions.\\


\section{Data reduction}\label{data_reduction}
In the following, we describe the data acquisition and data reduction process of the submillimetre 
lines observed with the Sub-Mm Array Receiver for Two frequencies (SMART) at the NANTEN2 telescope, 
and the Heterodyne Instrument for the Far-Infrared (HIFI) on board the Herschel satellite. 
We conclude with a summary of the main parameters characterizing our observations.\\

\subsection{Herschel-HIFI [CI]1-0, [CI]2-1, [NII], and [CII] observations}

With the Herschel-HIFI instrument, large-scale mapping of four submillimetre lines towards the Sgr A Complex was carried out  
in the frame of the  Herschel Guaranteed Time HEXGAL Key Program 
(P.I. R. G\"usten, MPIfR): \CILINEAIl\textrm{ }(OBS IDs 1342205504 and 1342205505), \CILINEAIIl\textrm{ }(OBS 
IDs 1342204733 and 1342204734), \NIIl\textrm{ }(OBS IDs 1342205892, 1342205893, 1342205894, 1342206599, 1342216664, 
and 1342216665), and \CIIl\textrm{ }(OBS ID 1342206647, 1342206648, 1342206649, 1342206650, 1342206651, and 1342206652). 
The HIFI Instrument is a double sideband (DSB) single  pixel receiver, with a total of seven mixer bands, 
ranging in frequency continuously from 480 GHz to 1250 GHz and from 1410 up to 1910 GHz. 
The mixing devices are SIS (superconductor-insulator-superconductor) mixers for bands 1 to 5 (4 GHz bandwidth) and 
HEB (hot-electron bolometer) mixers for bands 6 and 7 
(2.4 GHz bandwidth). Each band is in turn divided into two sub-bands, labelled with the letters 
``a'' and ``b'', each one associated with a different local oscillator (LO) chain. The backends used were 
two wide-band acousto-optical spectrometers (WBS), one for each measured vertical (V) and horizontal (H) polarization.
The spectral resolution of the WBS is 1.1 MHz and possesses a total bandwidth of 4 GHz \citep{degraauw2010}. 
The bands 1a, 3a, 6a, and 7b, correspond to the \CILINEAI, \CILINEAII, \NII, and \CII\textrm{ }observed atomic and ionic transitions, 
respectively. The pointing accuracy of the telescope was estimated to be $\sim$ 2'' and the calibration error in the measured 
absolute antenna temperatures ranges from 15\% in band 1a up to 25\% in band 7b \citep{roelfsema2012}.\\

\subsubsection{Observing strategy}

The observations were carried out in the equatorial coordinates reference frame and two observing modes were used for 
different frequencies: the \emph{On-The-Fly (OTF) Map With Position-Switch Reference} 
observing mode for the \CILINEAI\textrm{ }line, and the \emph{OTF Map With Load Chop And Position-Switch Reference} 
observing mode for the \CILINEAII, \NII, and \CII\textrm{ }lines. The choice between them is mainly driven by the Allan stability 
time \citep{allan1966,volker2008} of the receiver at a given frequency \citep{roelfsema2012}.\\

\begin{figure}
\centering
\includegraphics[angle=90,width=\hsize]{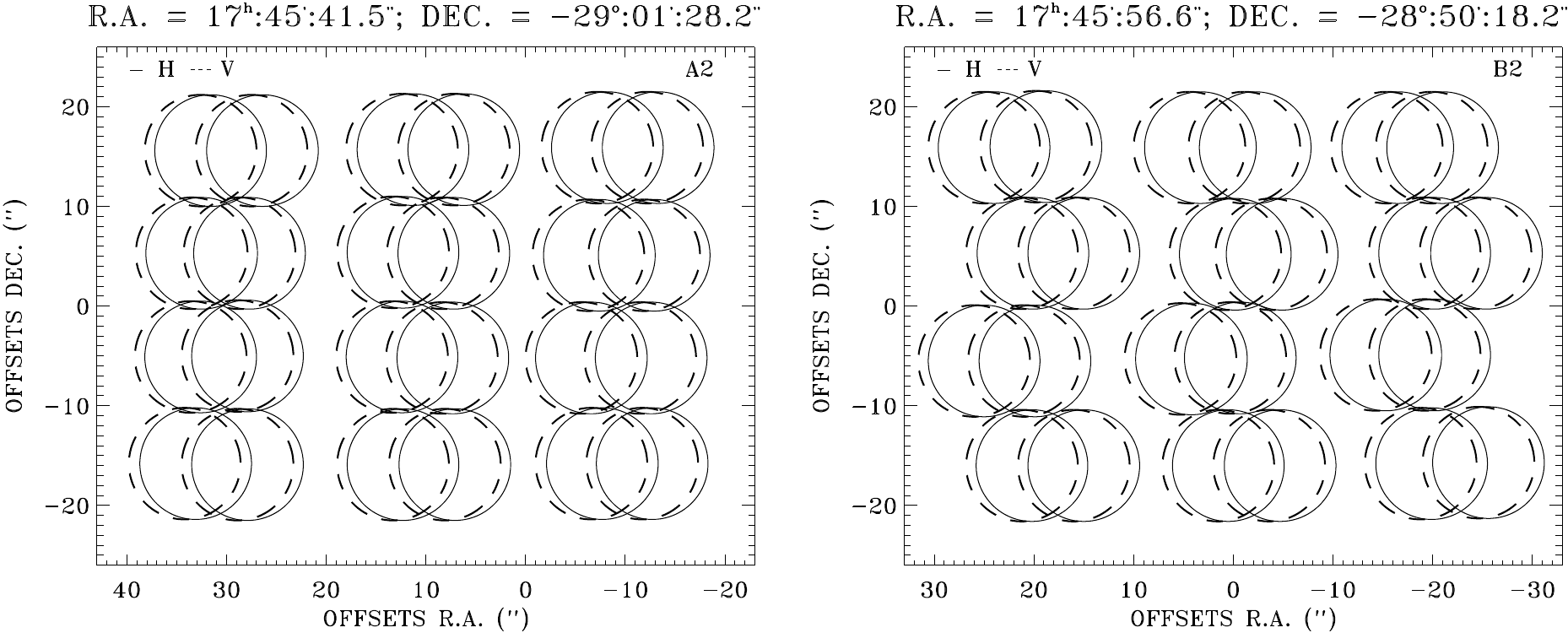}
\caption
{Examples of the \CII\textrm{ }observed grid in two different subregions of the entire observed map. The beam size at this frequency 
is represented by the circle's size. The H polarization (solid line) and V polarization (dashed line) spectra are shown. 
The slight misalignment between the beams is $-$1.0'' in the Y and 0.0'' in the Z focal plane coordinates of the HIFI
instrument. The \emph{OTF Map With Load Chop And Position-Switch Reference} observing mode used for the 
high frequency lines undersampled the observed maps.}
\label{fig_data_reduction:CII_sampling_example}
\end{figure}

The \emph{OTF Map With Load Chop And Position-Switch Reference} is very efficient in producing integration times on source that are 
shorter than the Allan stability times of the high frequency lines;  however, one disadvantage is that  it yields an irregular sampling pattern of the 
observed area leading to observed maps that are undersampled in the scanning direction given the beam size at the observed frequency.
An example of this is shown in Figure \ref{fig_data_reduction:CII_sampling_example} for the \CII\textrm{ }line. The figure shows the sampling 
pattern of two different subregions observed. The absolute equatorial coordinates of the central position (0,0) of each 
map are given at the top of each panel. The beam size at this frequency is 11.2'' and is represented by the plotted circles. Also, both 
polarizations, H (solid circle) and V (dashed circle), are shown. The beams of the two polarizations are shifted as a result of a small misalignment 
due to the construction process, measured to be $-$1.0'' in the Y and 0.0'' in the Z focal plane coordinates of the receiver \citep{roelfsema2012}. 
The angular separation between two consecutive spectra is 5.1'', while the separation for the spectra immediately before and after the chop 
to the internal load is 15.3'', larger than the beam size at this frequency. The separation of the OTF lines in the Declination is 10.5'', 
almost the size of the beam. From the figure it can be seen that the original measured map is undersampled in the scanning direction 
(Right Ascension). In the Declination direction, the map is beam sampled by design. The same sampling pattern is present in the 
\NII\textrm{ }and \CILINEAII\textrm{ }lines observations though the undersampled area is lower given the larger beam sizes.\\

\subsubsection{HIFI data reduction process}

\begin{figure}
   \centering
   \includegraphics[angle=-90,width=\hsize]{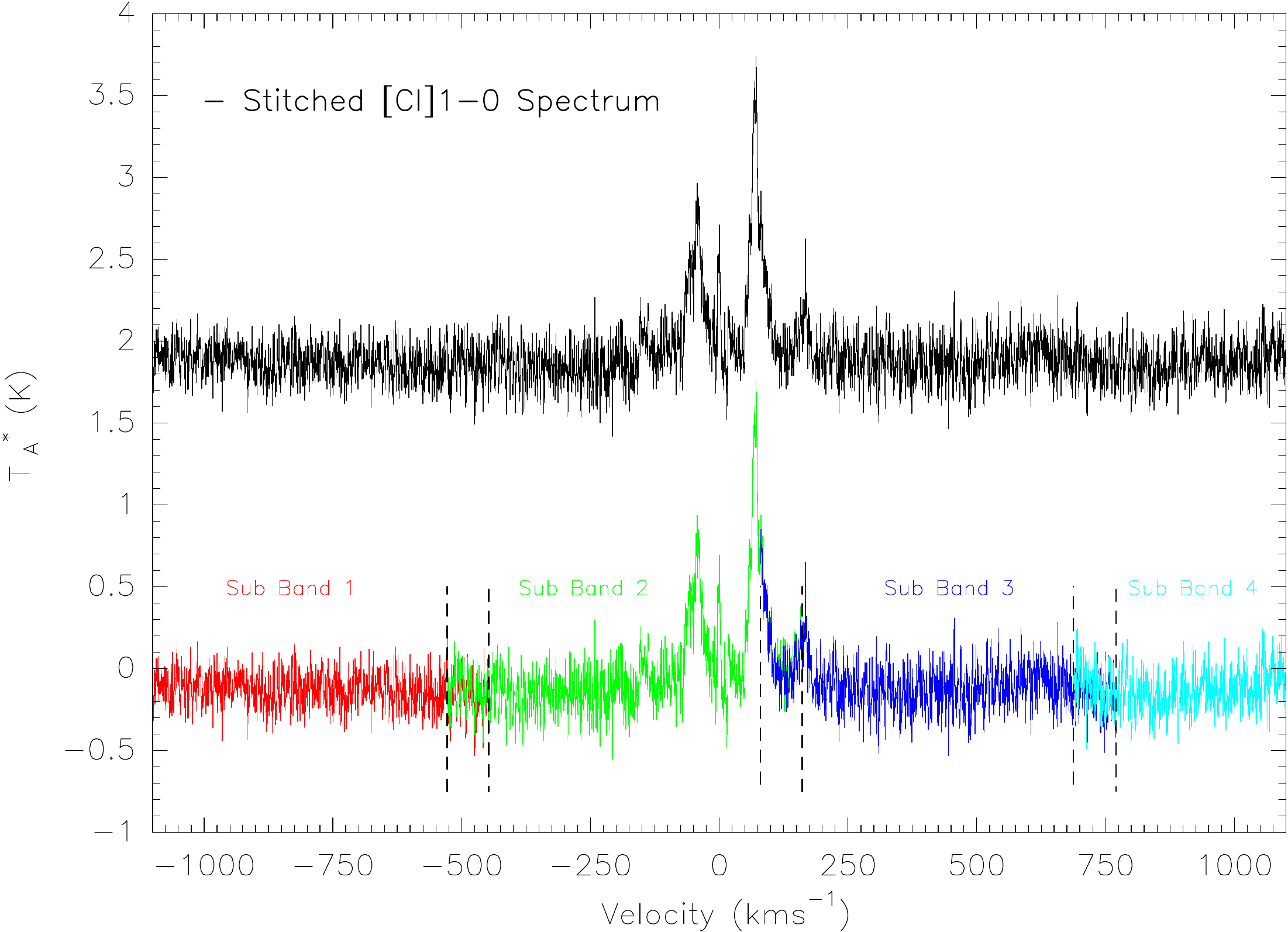}
   \caption{\CILINEAI\textrm{ }spectrum showing the 1 GHz WBS sub-bands before (colours) and after (black) the stitching procedure. 
The dashed vertical lines encompass the overlapping spectral region between consecutive sub-bands. After the stitching procedure, 
the full spectrum is used in the data reduction process. The displacement in the continuum level towards higher antenna 
temperatures of the end product spectrum is artificially introduced for display purposes}
\label{fig_data_reduction:subband_stitching}
\end{figure}

The data reduction software for all instruments on board the Herschel satellite is the \emph{Herschel Interactive Processing Environment 
(HIPE)} \citep{ott2010}. For each instrument a data processing pipeline processes the raw data counts into frequency and intensity calibrated data. 
This process is broken  into a number of discrete processing steps unique to each observing mode \citep{roelfsema2012}. The output from 
the pipeline at different steps can be examined, which is especially useful in dealing with situations in which the final output appears problematic.
For this work, the level2 data were taken. The first processing step after the standard level 2 pipeline is the ``Stitch'' Task (full name 
\mbox{\emph{herschel.ia.toolbox.spectrum.StitchSpectrumTask})}. This task takes the overlapping region among the sub-bands, 
calculates the average values in that range separately, brings them to the same continuum level, and glues the sub-bands together. 
An example of this procedure is shown in Figure \ref{fig_data_reduction:subband_stitching} for the observed \CILINEAI\textrm{ }line. 
The colours represent the four different 1 GHz WBS sub-bands. The velocity resolution at this frequency is 0.305 \KMS. The overlapping 
regions between sub-bands are enclosed between vertical dashed lines. The results of the stitching procedure are shown in the full 
spectrum (black line) on the same plot.\\

The raw spectra were exported to GILDAS compatible FITS (Flexible Image Transport System) files using the ``Hiclass'' task in HIPE 
 to further continue the data reduction process in the CLASS module of GILDAS software (\url{https://www.iram.fr/IRAMFR/GILDAS/}). 
In the case of the Sgr A Complex the emission lines cover a very wide LSR velocity range, between $-$200 \KMS\textrm{ }and $+$200 \KMS. Since spectral resolution 
of the HIFI receiver is very high, we resampled all spectra to 1 \KMS\textrm{ }spectral resolution. This ensures that emission lines are 
spectrally resolved,  at the same time increasing  the signal-to-noise ratio. For the baseline subtraction, polynomials 
of order  1, 2, and  3 were used. Spectra that have anomalies in their baselines can usually be filtered out by setting a threshold to 
the rms temperature (T$_{A,rms}^{*}$) tolerance, but given the excellent stability of the HIFI receiver in bands 1a and 3a, 
none of the spectra were discarded.\\

\subsubsection{Special cases}\label{special_cases}

The baseline fitting procedure for the \NII\textrm{ }and \CII\textrm{ }lines in bands 6a and 7b, respectively, requires a slightly 
different approach. The HEB mixers in bands 6a and 7b show a reflection in the electrical amplification chain, 
which leads to distortion of the baselines in the spectra measured at these frequencies \citep{higgins2011}.\\

In band 6a, the spectra show a baseline oscillation at one edge of the bandpass, between $+$94 \KMS\textrm{ }and $+$170 \KMS, 
in  the  V and the H polarizations. This perturbation is shown in Figure \ref{fig_data_reduction:NII_lo_problem} (see Appendix \ref{appendixB}) 
as a red line in the same LSR velocity range. The spectrum shown is the average spectrum of the H polarization over the whole data set.
Unfortunately, given the very wide velocity dispersion of the line, this perturbation affects regions of the bandpass
where emission is present. The origin of this anomaly is related to a combination of problems: the diplexer coupling at this 
frequency, the electrical standing waves in the HEB mixers, and the IF (intermediate frequency) frequency dependence sensitivity of the 
HEB mixers that  are more sensitive to lower IF frequencies \citep{higgins2011}. In order to minimize the impact of this oscillation on the final data set, 
the average T$_{A}^{*}$ value between $+$94 \KMS\textrm{ }and $+$170 \KMS\textrm{ }was calculated and assigned to the measured antenna temperature 
before subtracting the baseline from the spectra. The result of this procedure is shown as a black dashed line in the 
same figure. We investigated the effect of a polynomial of order 0 and 1 subtracted from the spectra and the best results were 
obtained with the latter. In this way, the usable LSR velocity range of this data set 
is from $-$200 \KMS\textrm{ }to $+$94 \KMS\textrm{ }only.\\

\begin{figure}
   \centering
   \includegraphics[angle=-90, width=\hsize]{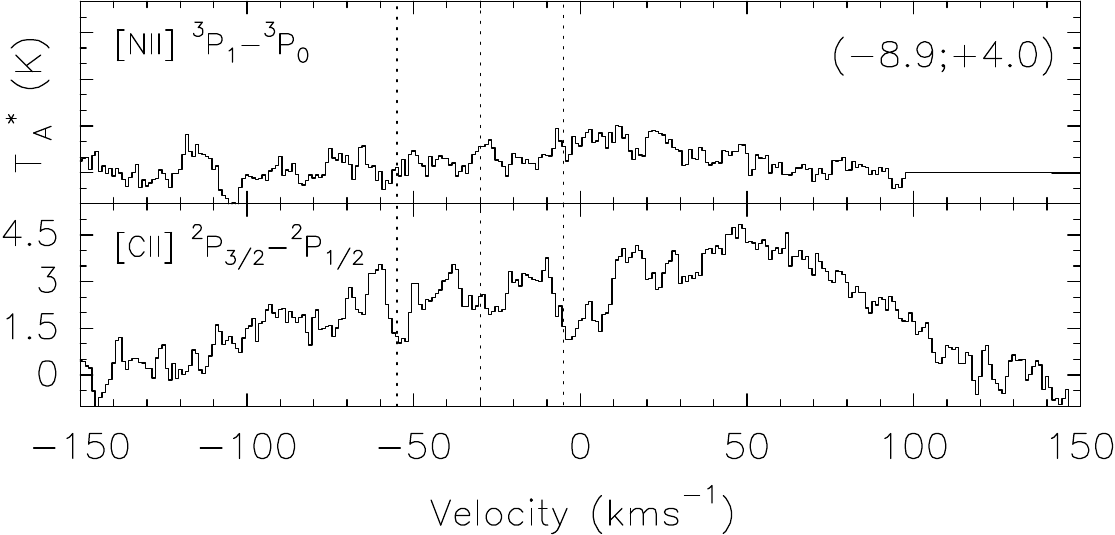}
   \caption{\NII\textrm{ }and \CII\textrm{}spectra at position $\Delta \alpha(J2000)=$ $-$8.9'', $\Delta \delta(J2000)=$ $+$4.0'' after 
   the filtering process explained in Section \ref{special_cases} and after the final data cube was created. The straight line at LSR velocities 
   $>$ $+$94 \KMS\textrm{ }in the \NII\textrm{ }spectrum is the result of the method utilized in removing the baseline distortion (see text).}
\label{fig_data_reduction:examples_lo_problem}
\end{figure}

In band 7b, the V polarization HEB mixer suffers from a too high LO power, which yields 
an impedance mismatch between the mixer and the first low noise amplifier (LNA) as was shown by \citet{higgins2011}. Since both 
polarizations share the same LO chain, optimizing the LO power for one of the polarizations unavoidably causes the LO power of the other polarization to be
 either  too low (in the case of the H polarization HEB mixer) or too high (V polarization HEB mixer). In the 
Herschel-HIFI observation scheme, it was decided to operate the H polarization HEB mixer at the optimum level, namely, 
at a mixer current of 0.045 mA, and the V polarization mixer current was left to follow \citep{higgins2011}. As a consequence, 
some of the V polarization spectra suffer from large distortions of their baselines. An example of the V polarization spectra is 
shown in Figure \ref{fig_data_reduction:CII_lo_problem} (see Appendix \ref{appendixB}). The figure shows the average spectrum of one OTF line. 
We set a T$_{A,rms}^{*}$ $=$ 2.3 K threshold to filter out the bad spectra from the V polarization data set. For comparison with Figures 
\ref{fig_data_reduction:NII_lo_problem} and \ref{fig_data_reduction:CII_lo_problem}, single \NII\textrm{ }and \CII\textrm{ }spectra at position 
$\Delta \alpha(J2000)=$ $-$8.9'', $\Delta \delta(J2000)=$ $+$4.0'' (the closest position to the centre of our maps) after the final data 
cubes were created are shown in Figure \ref{fig_data_reduction:examples_lo_problem}. The  straight line at LSR velocities $>$ $+$94 
\KMS\textrm{ }in the \NII\textrm{ }spectrum is the result of the interpolation process applied to that data set.\\

\subsubsection{Emission contamination in the reference position}

\begin{figure}
\centering
\includegraphics[angle=-90, width=\hsize]{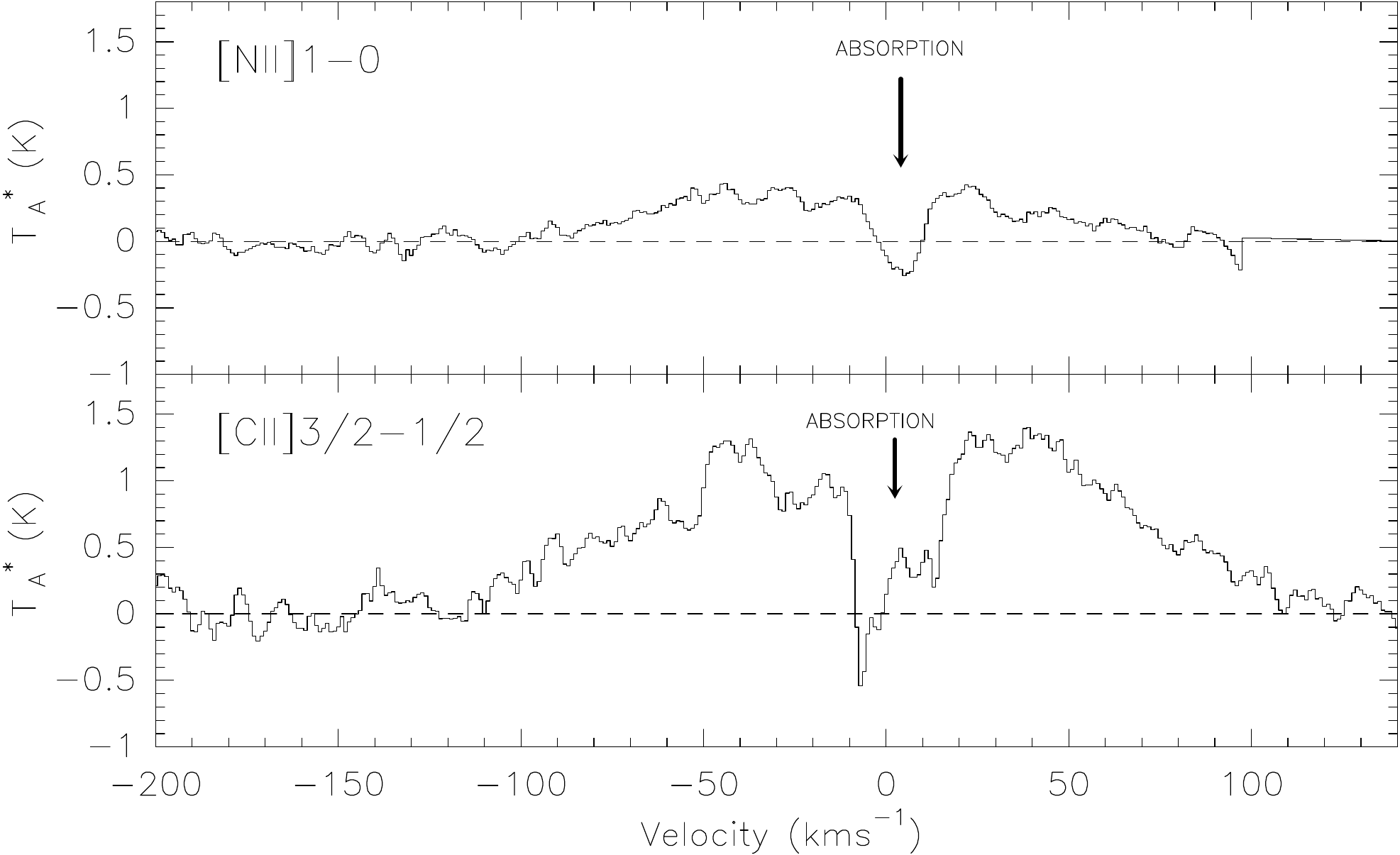}
\caption{Average spectra of the Herschel-HIFI \NII\textrm{ }and \CII\textrm{ }observations. The arrows show the 
positions of the absorption features present in the data due to the contamination of the reference position (SKY)
with line emission.}
\label{app:absorptions}
\end{figure}

The reference position of the Herschel-HIFI observations is located at $\alpha(J2000) =$ 17$^{h}$\textrm{ }44'\textrm{ }33'', 
$\delta(J2000) = -$28\DEG\textrm{ }52'\textrm{ }08'' in equatorial coordinates for all four lines observed. The selection of 
this position was based on a compromise between the absence of polycyclic aromatic hydrocarbon emission detected by the SPIRIT III infrared telescope at 
8.3 $\mu m$ as part of  the Midcourse Space Experiment (MSX) Galactic Plane Survey, and a reference position not far  
from the astronomical target. Unfortunately, all four lines observed with the Herschel-HIFI receiver suffer from emission 
contamination in the reference position. Figure \ref{app:absorptions} shows the average spectrum of the \NII\textrm{ }and 
\CII\textrm{ }lines over the entire observed region.  A massive absorption feature can be seen in both spectra, 
around  0 \KMS, going well below the 0 K antenna temperature in the baseline subtracted spectra.\\

Absorption features are expected to be found when observing towards the Galactic Centre due
to the presence of foreground material (such as spiral arms) along the l.o.s. \citep{linke1981}. In order to 
identify whether the absorption features seen in Figure \ref{app:absorptions} are the result of a contaminated 
reference position or of the foreground absorbing material,  a calibrated spectrum
of the reference position is needed in order to assess whether there is emission there. The most common way to 
recover the emission of a contaminated reference position is to select an area of the observed
region that contains no emission, usually towards the edges of the maps. From the selected spectra, an average spectrum 
with lower noise temperature than the rest of the map can be created. After that, the observed absorption features are 
inverted and added to all the spectra in the map. Unfortunately, this approach cannot be used in the case of the 
Herschel-HIFI observations. The emission distribution is very extended in all lines, so there is no position free 
of emission in the map. Therefore, a different approach must be used.\\

Given the observing mode used in some of the Herschel-HIFI observed data sets, it is possible to
reconstruct the reference position spectrum for each of the OTF lines observed. The calibration of the measured reference position 
is possible for all lines observed in the \emph{OTF Map With Load Chop And Position-Switch Reference} 
observing mode such as \CILINEAII, \NII, and \CII. This observing mode chops to internal loads immediately before taking a spectrum. 
The chop to internal loads is used to determine the bandpass of the receiver and is done for the on source position as well as for the 
reference position. In this way, the standard data reduction pipeline of the Herschel-HIFI observations can be applied to convert the
measured counts at the reference position into fully calibrated reference spectra in the $T_{A}^{*}$ temperature scale. Nonetheless, 
there are telescope related artefacts, such as standing waves produced by the difference in light path when chopping to the internals loads, 
that are not removed from the SKY spectra during the standard calibration process. In this sense, the reference spectra are calibrated properly 
in temperature scale, but contain standing waves, whose shapes depend on the utilized spectral band and have to be removed in order to recover 
the line emission. This procedure is explained in Section \ref{model_ref_position}. The corresponding amplitudes of these standing waves are of 
the order of $\sim$ 0.5 K for the \CII\textrm{ }and \NII\textrm{ }lines, and $\sim$ 1 K for the \CILINEAII\textrm{ }line.\\

\subsubsection{Model of the reference position emission}\label{model_ref_position}

In order to account for the subtracted reference emission from the data, we modelled the emission at the 
reference position and added it back to each individual spectrum in a series of steps:  
(a) We calibrated each observed reference position producing a calibrated spectrum in $T_{A}^{*}$ temperature scale; 
(b) we fit a 0-order polynomial to the spectra and selected only the ones with the lowest noise. 
Thus, an average spectrum with much lower noise than the spectra in the map is created. 
The selected spectra used to create the average spectrum are $\sim$ 100 for the \CILINEAII\textrm{ }line and around $\sim$ 200 for 
the \NII\textrm{ }and \CII\textrm{ }lines; (c) we removed only well-defined standing waves identified in a Fourier transform of the average spectrum.
This was possible because all standing waves have a consistent structure in all spectra, so adding them  increases
the signal-to-noise ratio but preserves their structure. Well-defined standing waves appear as very narrow spikes in the time domain, 
so it is, in principle, possible to remove them by interpolating the intensity just before and after the spike without affecting the
true emission structure of the spectrum. Ideally, after doing this, the velocity structure of the line should be compared with other 
observations -- obtained by independent observations -- of similar tracers at the same position  in order to avoid introducing artificial 
structures to the data (see below); (d) a low-order baseline was fitted and subtracted from the resulting average spectrum 
where the emission was already clearly visible; (e) we modelled the emission by fitting up to five independent Gaussians 
to the average spectrum creating an {emission model} that was added to each spectrum in the observed map.
This procedure effectively recovers the emission contamination in the reference position and avoids adding extra noise to the 
data apart from the uncertainties in the fitted Gaussian parameters.\\

We checked that no artificial emission was created during the removal of
the standing waves by trying different interpolations and checking the variations in the results. Only when 
the standing wave was very well defined as a narrow intensity peak in the time domain and its radial velocity coincided with 
the radial velocity corresponding to the absorption features in the average spectrum of the corresponding line 
(as shown in Figure \ref{app:absorptions}) was the resulting spectrum  used to create the emission model. 
The difference in the measured positions of the reference spectra is of the order of the absolute pointing 
error (APE) of the telescope $\sim$ 2'' \citep{roelfsema2012}, so it is assumed that all reference spectra 
were measured at the same position on the sky.\\ 

\begin{figure}
\centering
\includegraphics[angle=-90, width=\hsize]{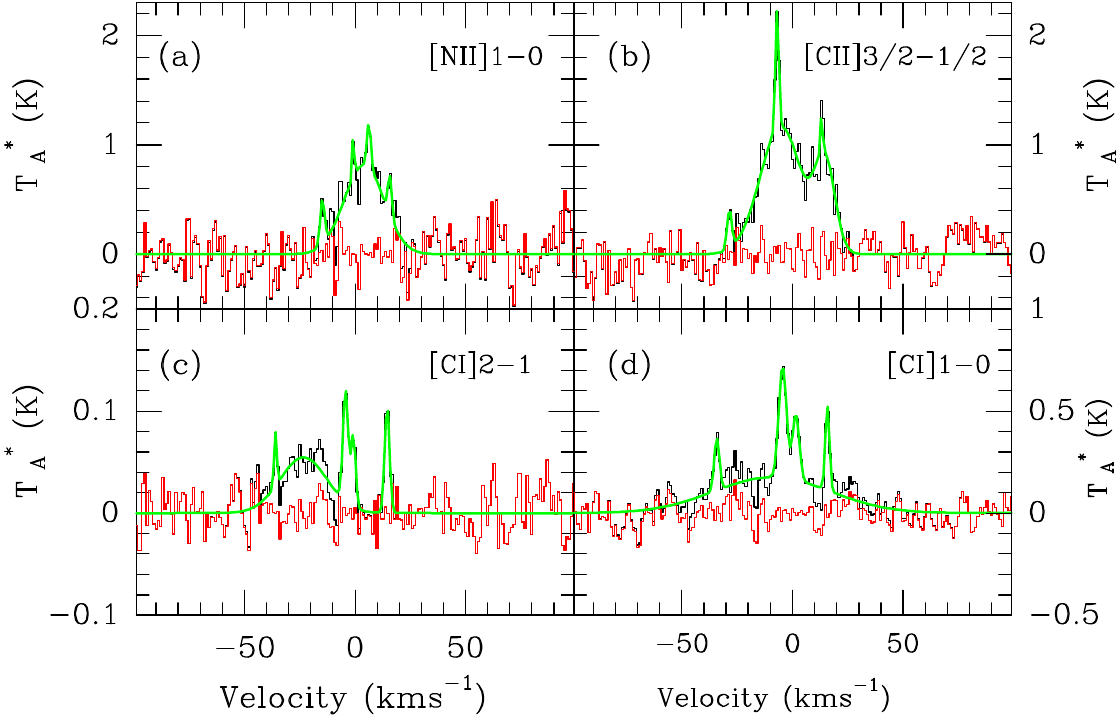}
\caption{Panels (a), (b), (c), and (d) show the reference position calibrated average spectrum (black), the fitted emission model  
(green), and the residual after the subtraction of the fitted model from the reference position emission (red), 
for the \NII, \CII, \CILINEAII, and \CILINEAI\textrm{ }observations, respectively.}
\label{app:herschel_models}
\end{figure}

The results of the procedure are shown in Figure \ref{app:herschel_models}. The black line 
in panels (a), (b), and (c) represent the average reference spectrum for the \NII, \CII\textrm{ }, and \CILINEAII\textrm{ }lines, 
respectively (see Section \ref{appendix:CI10_herschel} for \CILINEAI). The noise in each spectrum is $\sigma =$ 0.196 K for the \NII\textrm{ }line, 
$\sigma =$ 0.173 K for the \CII\textrm{ }line, and $\sigma =$ 0.022 K for the \CILINEAII\textrm{ }line.
In all three cases, the noise of the average reference spectrum is more than 10 times lower than the typical noise
in the map. The green line is the emission model constructed from the fitting of up to
five independent Gaussians to the observed emission. We reproduce all observed emission above a $\sim$ 
2.5$\sigma$ noise level. The quality of the model in reproducing the emission can be assessed from the 
residual spectra (red line) showing the difference between the observed spectrum and the fitted model across the whole velocity range. 
All three models reproduce  the emission contained in the average reference spectrum very well. {The  \CILINEAI\textrm{ }profile 
in panel (d), obtained by independent measurements of the reference position (with $\sigma =$ 0.064 K, see Section \ref{appendix:CI10_herschel}),
shows almost exactly the same velocity structure as the \CILINEAII\textrm{ }profile in panel (c), and is also
similar to the \CII\textrm{ }profile in panel (b); this strongly supports our approach to determining the emission structure
at the reference position.}\\
 
\begin{figure}
\centering
\includegraphics[angle=-90, width=\hsize]{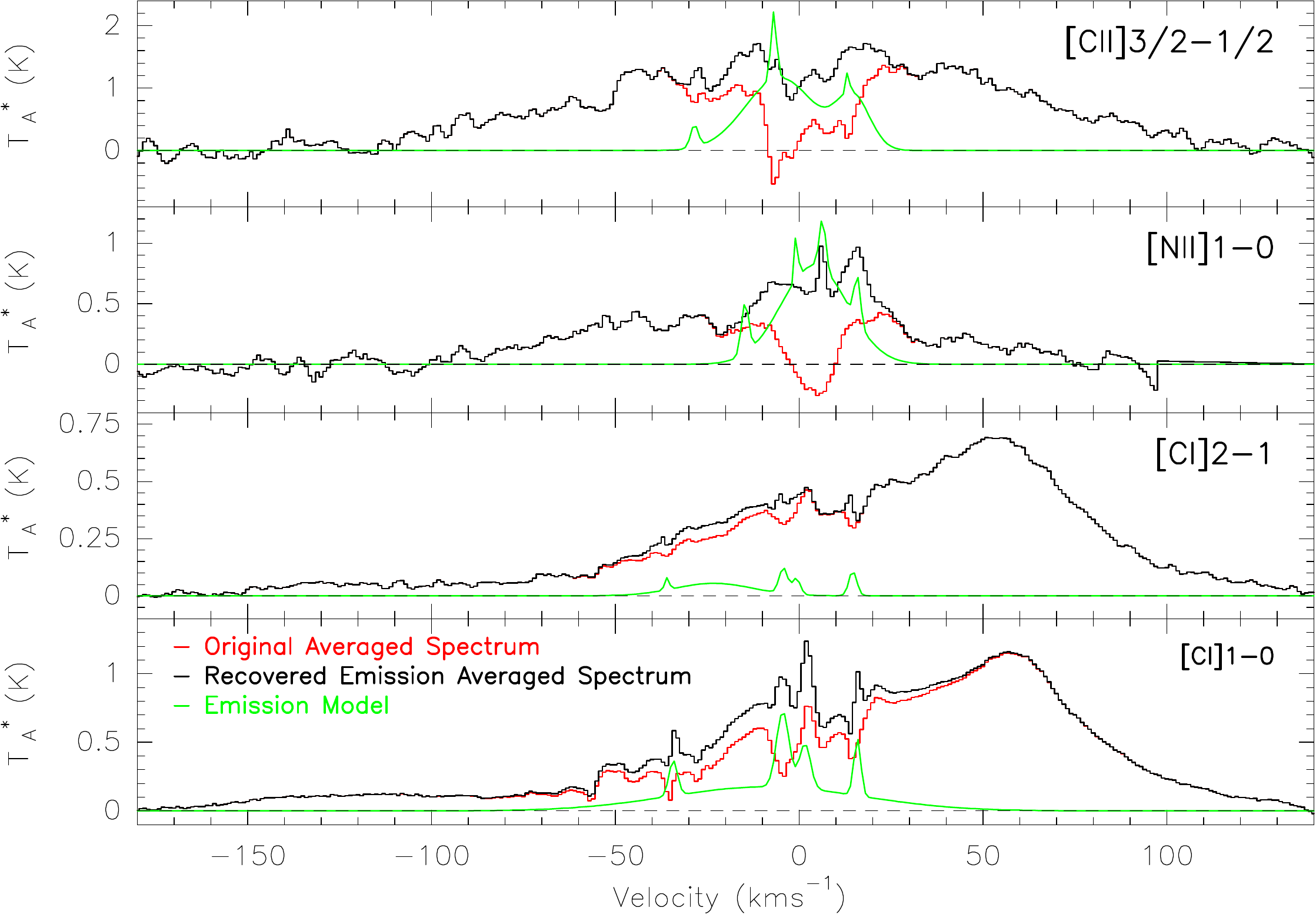}
\caption{From top to bottom, the \CII, \NII,  \CILINEAII, and \CILINEAI\textrm{ }average spectra before (red) and after (black) the
addition of the reference position emission model (green) to each individual spectrum are shown. The absorption
features due to the contamination of the reference position in all four lines are fully recovered.}
\label{app:herschel_results}
\end{figure}

We also compared the average spectrum before and after adding the emission model
to each individual spectrum (see  Figure \ref{app:herschel_results}). From top to bottom, the models of the \NII, \CII\textrm{ }, and \CILINEAII\textrm{ }emission 
in the reference position recover all the most important absorption features shown in Figure \ref{app:absorptions}. 
In the case of the \NII\textrm{ }line, the emission in the reference position is much stronger than the average emission 
over the map, so the model shape stands out more clearly than in the other lines, as seen in the intensity 
peak around $+$5 \KMS. After the recovery of the emission at the reference position has been added back to the data sets, 
we assume that only real physical absorption features along the l.o.s. are to be found.\\

\subsubsection{ \CILINEAI\textrm{ }reference emission}\label{appendix:CI10_herschel}

In the case of the \CILINEAI\textrm{ }line, the \emph{OTF Map With Position-Switch Reference} observing mode was used to carry out 
the observations so we adopted a different approach to check for emission contamination in the reference position. 
In this observing mode, given the stability of the receiver, no chop to the internal loads is needed before each
ON  read out. Instead, a load measurement was done after each OTF line. In principle, it is possible to obtain as many fully
calibrated reference spectra as there are OTF lines by using the load measurement and reference position integration closest in
time. Nonetheless, this approach requires a major modification of the standard HIFI data reduction pipeline. More importantly, 
we chose a different method than the one mentioned in the previous section in order to have independent measurement
of the \CILINEAI\textrm{ }profile at the reference position, so the quality of the method used for the other lines can be 
assessed (see  previous section). We used the NANTEN2/SMART telescope to re-observe the Herschel-HIFI 
reference position against a reference position that was further away. We observed a 30''$\times$30''  map, centred at the reference 
position, with 10'' spacing in both directions. Given the better spatial resolution of the 4 m NANTEN2/SMART telescope, 
we smoothed the map to a resolution equivalent to the beam size (FWHM) of the Herschel-HIFI telescope at this frequency (43.1''). 
The average spectrum of the map (Fig. \ref{app:herschel_models}, panel (d)) has a rms noise of $\sigma =$ 0.064 K in 
the $T_{A}^{*}$ antenna temperature scale, which is lower than the rms noise of all Herschel-HIFI spectra measured at this frequency. 
The emission model (green line) reproduces  the main features in the spectrum very well, and yields a residual spectrum (red line) 
consistent with pure noise.\\ 

In order to recover the emission subtracted from the ON spectra, we added the modelled emission channel by channel to 
the Herschel-HIFI \CILINEAI\textrm{ }spectra. We assume by this procedure that both telescopes measure the same antenna temperatures 
in the $T_{A}^{*}$ scale, an assumption that is fully supported by the analysis in Section \ref{appendix:temp_scale}. In Figure \ref{app:herschel_results}, 
a comparison between the average spectrum, before and after recovering the emission subtracted from the on-source spectra, is shown. 
The remaining absorption features in the average spectrum are due to foreground material along the l.o.s.\\

After the baseline subtraction and OFF emission recovery processes were performed, all data sets were brought to position-velocity data cubes 
using the Grenoble Graphic (GREG) software, also part of the GILDAS package. The spatial resolution of the \CILINEAI\textrm{ }and \CILINEAII\textrm{ }data cubes   
(46'' and 28'', respectively) is slightly lower  ($\sim$ 6\%) than the corresponding beam sizes of the observations (see Table \ref{tab:data_summary}).
This is due to the Gaussian convolution kernel in GREG, which is  one-third the size of the telescope beam width. For the case of the \NII\textrm{ }and \CII\textrm{ }lines, 
 we set the convolution kernel size so the data match the 46'' spatial resolution of the \CILINEAI\textrm{ }data cube, improving the signal-to-noise ratio
of the observations in the processs. The spatial noise distribution of the Herschel-HIFI data sets is shown in the panels of Figure  
\ref{fig_data_reduction:herschel_noise_distributions}, together with a histogram of the T$_{A,rms}^{*}$ values (upper panels). The Gaussian profiles 
fitted to the T$_{A,rms}^{*}$ histograms are used only as a convenient approximation of the T$_{A,rms}^{*}$ true distribution to quantify the characteristic 
noise of each data set. A more complex chi-squared distribution would reproduce the T$_{A,rms}^{*}$ distribution more accurately \citep{goldsmith2002}, 
but such an analysis is outside the scope of the present work.\\

\begin{figure}
 \begin{minipage}{\hsize}
   \centering
   \includegraphics[angle=0,width=\hsize]{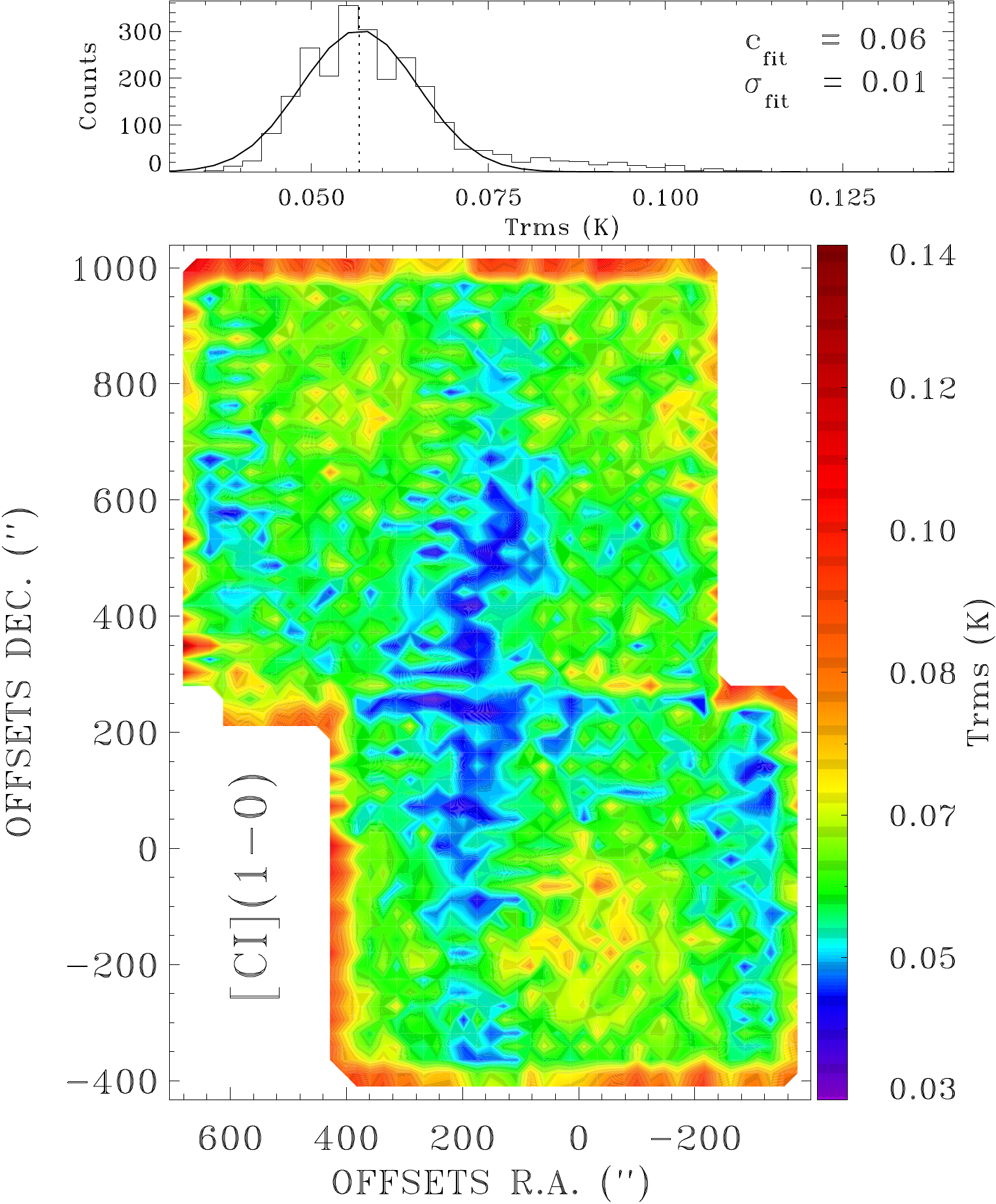}
 \end{minipage}
 \begin{minipage}{\hsize}
   \centering
   \includegraphics[angle=0,width=\hsize]{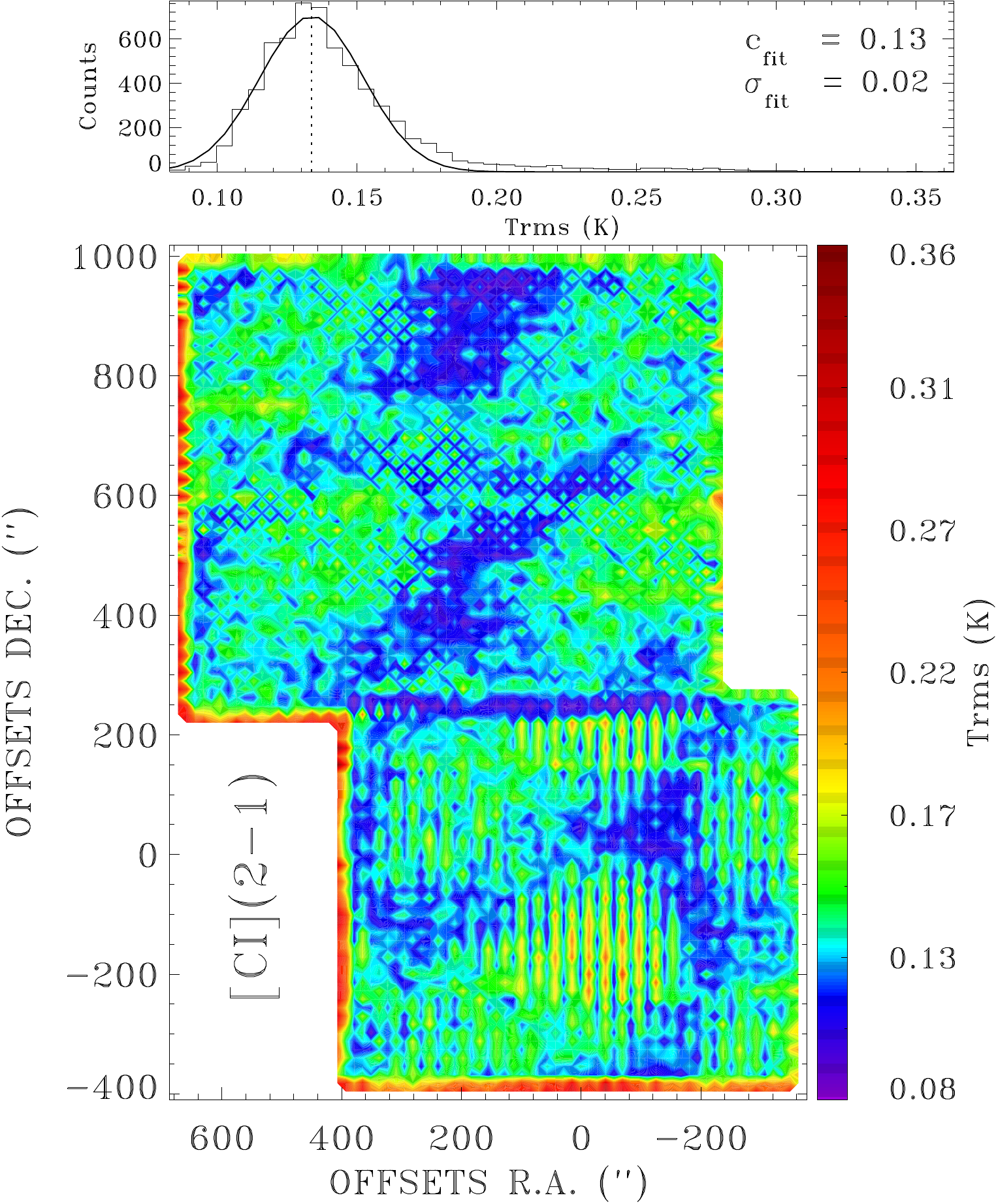}
 \end{minipage}
\end{figure}
\begin{figure}
 \begin{minipage}{\hsize}
   \centering
   \includegraphics[angle=0,width=\hsize]{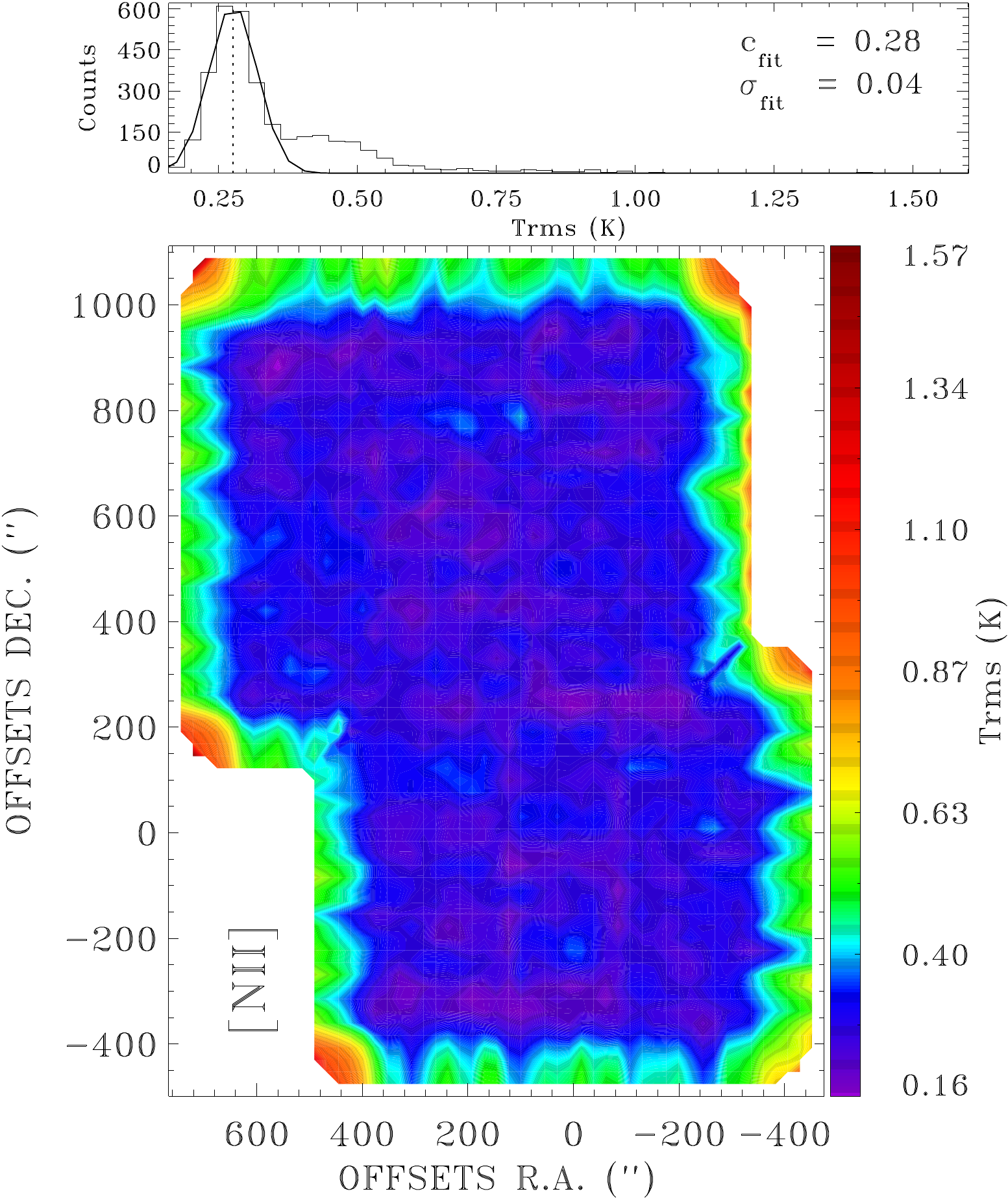}
 \end{minipage}
 \begin{minipage}{\hsize}
   \centering
   \includegraphics[angle=0,width=\hsize]{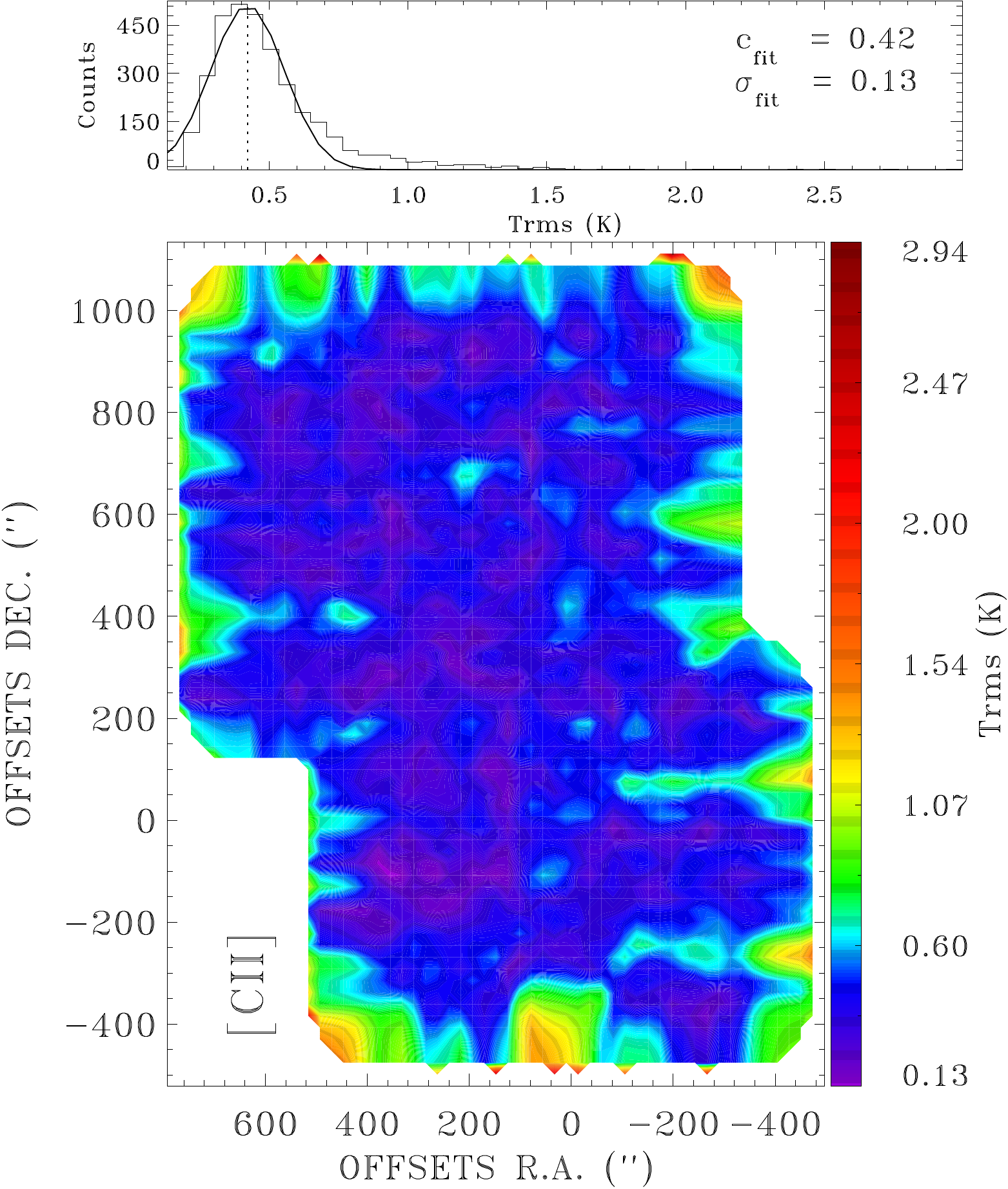}
 \end{minipage}
  \caption{\CILINEAI, \CILINEAII, \NII, and \CII\textrm{ }T$_{A,rms}^{*}$ noise distributions. For each panel, the spatial noise 
distribution as well as the histogram of the noise distribution are shown. The noise distribution spatial maps are centred at the 
$\alpha(J2000) =$ 17$^{h}$\textrm{ }45'\textrm{ }39.9'', $\delta(J2000) = -$29\DEG\textrm{ }00'\textrm{ }28.1'' position. 
From the spatial maps, the noise is higher at the edges of the maps, as expected given the small number of spectra at the borders with which 
to calculate the resampled spectrum. From the histograms, a Gaussian fit (solid curve) shows the typical noise of each map $C_{fit}$ 
(Gauss centre shown as a vertical dashed line) and the standard deviation of the distributions $\sigma$$_{fit}$.}
\label{fig_data_reduction:herschel_noise_distributions}
\end{figure}


\subsection{NANTEN2/SMART CO(4-3) observations}

As part of the NANTEN2/SMART Central Molecular Zone Survey, the \COLINEAI\textrm{ }line was observed towards the Sgr A Complex. 
Other submillimetre observations such as the \CILINEAI\textrm{ }line were carried out to obtain a cross-calibration between the 
Herschel-HIFI and NANTEN2/SMART observations as shown and discussed in Section \ref{crosscal}. The SMART instrument
\citep{urs2003} is a DBS 2$\times$8 pixel heterodyne receiver operating at 460 GHz and 810 GHz frequencies simultaneously. The 
tunable range of the two frequencies spans from 435 GHz to 495 GHz for the 460 GHz channels and from 795 GHz to 880 GHz for the 
810 GHZ channels, with an intermediate frequency ($\nu_{IF}$) of 4 GHz and 1.5 GHz, respectively. The mixing devices are 4 GHz 
bandwidth SIS mixers for both channels. During the data acquisition campaign, two different backends were used: Acousto Optical Spectrometers (AOSs) \citep{horn1999} 
and eXtended bandwidth Fast Fourier Transform Spectrometers XFFTSs \citep{klein2012}. The AOSs were used from the beginning of the 
data acquisition campaign until August 2013. They have a 1 GHz bandwidth with a 1.04 MHz spectral resolution corresponding to 
0.68 \KMS\textrm{ }at 460 GHz. About 75\% of the data were observed using the AOSs as the backend before switching to the new XFFTSs. 
The XFFTSs were installed in September 2013. They posses a total bandwidth of 2.5 GHz and a spectral resolution of 88.5 KHz, 
corresponding to 0.05 \KMS\textrm{ }at 460 GHz. With both backends, we measured essentially the same antenna temperatures within 
calibration errors, as expected. The typical pointing error of the 4 m NANTEN2/SMART telescope was within $\sim$ 10'' and the absolute calibration 
error for the 460 - 490 GHz range, including uncertainties in beam efficiencies \citep{simon2007} and in the atmospheric calibration (see Section 
\ref{SMAT_data_reduction}), is estimated to be around 20\% \citep{pineda2008,pineda2012}.\\

\subsubsection{Observing strategy}

The  observations of the \COLINEAI\textrm{ }line were carried out in the Galactic coordinates reference frame using the OTF observing mode. 
The basic sampling unit of the SMART receiver, called the {footprint,} is a 340''$\times$170'' map that is Nyquist sampled at 810 GHz and 
oversampled at 460 GHz. This spacing increases the signal-to-noise ratio of the lower frequency line after spatial resampling to the angular 
resolution consistent with the beam size. After each OTF line an OFF position measurement was performed, and a calibration load measurement was 
done every two OFF measurements.\\

The identification of a clean reference position for the observations is an enormous problem for large-scale surveys of the Galactic Centre. 
As previously mentioned, all Herschel-HIFI observed lines suffered from emission contamination in the reference position that could be 
recovered largely because the receiver has only one pixel and no atmosphere had to be taken into account for the data calibration process. 
In the case of the NANTEN2/SMART observations, the picture is much more complex since during the OFF measurement each of the eight pixels 
at 460 GHz points through the atmosphere at a different position on the sky, typically separated by 85''. This means that not only 
a single clean position has to be found to be used as a reference but rather an area containing the observed positions in the reference measurement. 
Therefore, and based on the CO(2-1) observations of 
\citet{sawada2001}, the $l = +$0\DEG.497, $b = +$0\DEG.980 position was chosen. This position is located $\sim$ 1\DEG\textrm{ }away from 
the Galactic Centre. In order to check whether there is still emission contamination, despite  the great distance to the reference position,  
we observed the reference position against a more distant reference position that had been previously checked for emission contamination and found to be clean. 
The observations showed no emission within the footprint containing all eight pixels, at least down to a 3$\times$T$_{A,rms}^{*}$ significance level. Unfortunately, 
selecting a position far away from the astronomical source also has  some  drawbacks, especially in terms of the receiver stability 
and airmass between on-source and reference position measurements. This causes   baseline distortions to appear in some of the 
calibrated spectra.\\

\subsubsection{SMART data reduction process}\label{SMAT_data_reduction}

\begin{figure}
\centering
\includegraphics[angle=0,width=\hsize]{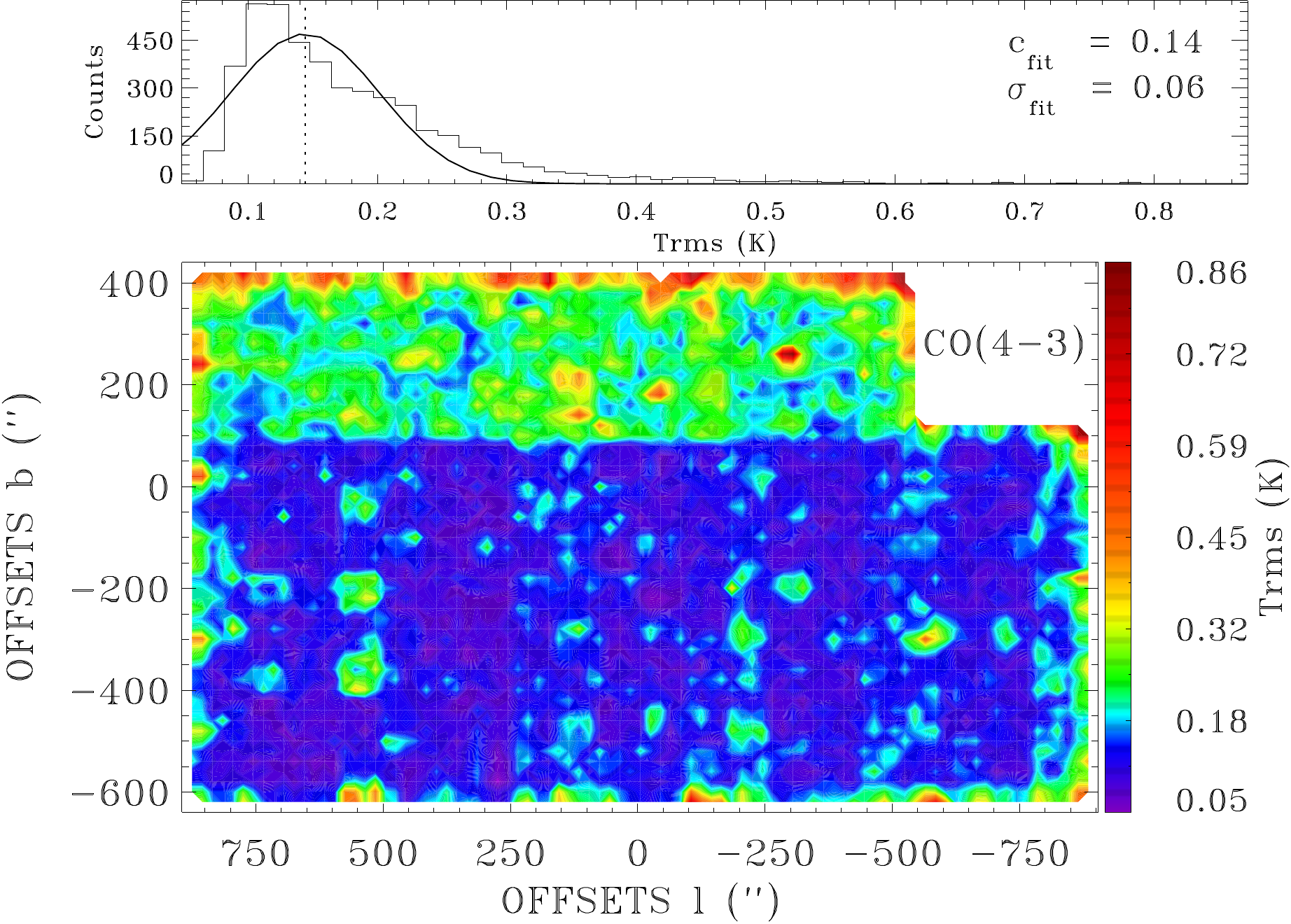}
\caption{\COLINEAI\textrm{ }T$_{A,rms}^{*}$ noise distributions. The spatial noise distribution and the histogram of the noise 
distribution are shown. The noise distribution spatial map is centred at  l $=$ 0\DEG, b $=$ 0\DEG\textrm{}. 
From the spatial map, the noise is higher at the edges of the map, as expected given the small number of spectra at the borders with which 
to calculate the resampled spectrum. From the histogram, a Gaussian fit (solid curve) shows the typical noise of the map $C_{fit}$ 
(Gauss centre shown as a vertical dashed line) and the standard deviation of the distribution $\sigma$$_{fit}$.}
\label{fig_data_reduction:nanten_noise_distributions}
\end{figure}

In order to calibrate the observations, the atmospheric transmission as a function of frequency $t_{a}(\tau(\nu))$ for each measured spectrum is needed. 
In order to determine $t_{a}(\tau(\nu))$ in each side band, additional information of the atmosphere's behaviour 
is obtained from the atmospheric model (ATM) developed by \citet{pardo2001}, adapted for the telescope's altitude at 4865 m 
above sea level. From the model and the observations, the amount of precipitable water vapour ($m_{H_{2}O}$) was fitted,  the 
atmospheric opacity $\tau_{\nu} \propto m_{H_{2}O}$ was obtained, and the $t_{a}(\tau(\nu))$ value was derived \citep{guan2012}. After all observations 
were calibrated, the data were further reduced with the CLASS package where the spectra were brought to the forward 
beam antenna temperature scale T$_{A}^{*}$ and low order (0 to 3) polynomials were fitted and subtracted from them (see Section \ref{crosscal} for an 
explanation on the selected antenna temperature scale). We only kept spectra with rms noise T$_{A,rms}^{*}$ $<$ 2 K, signal-sideband opacity 
$\tau_{s} <$ 0.7 (which implies an attenuation of the sky signal by the atmosphere of about 50\%), and system temperature $T_{sys} <$ 800 K. 
Any spectrum with at least one of the three parameters above the corresponding thresholds was discarded. The final data cube was created 
with a spatial resolution (40'') slightly lower than the beam size (37.4'') at this frequency. The spatial noise distribution of the 
NANTEN2/SMART data is shown in Figure \ref{fig_data_reduction:nanten_noise_distributions}.\\

\subsection{Antenna temperature scale and Herschel-HIFI -- NANTEN2/SMART cross-calibration}

\subsubsection{Temperature scale}\label{appendix:temp_scale}

In order to choose a common absolute antenna temperature scale for combining the NANTEN2/SMART and Herschel-HIFI observations, we examine the 
emission distribution of the \mbox{Sgr A} region and some aspects of the antenna patterns of both telescopes. Table \ref{tab:data_summary} contains all the
relevant parameters used to define the absolute temperature scales. The spatial extension of the submillimetre emission in the Sgr A region,
as can be seen in \citet{garcia2015}, is very large compared to the beam sizes and side lobes in all lines included in the present work. 
For the Herschel satellite, \citet{mueller2014} improved the analysis of the Herschel  beam pattern done by \citet{roelfsema2012} 
using all Mars observations performed during the  Herschel  operations together with an optical model of the telescope, which includes
obscuration, truncation, and measured wave-front errors. They derived more accurate estimates for the beam sizes ($\Theta_{FWHM}$) and
main beam efficiencies ($B_{eff}$) for each measured polarization,  while in Table \ref{tab:data_summary} the quoted 
main beam efficiencies are the averaged values of the two  polarizations. We kept the beam sizes estimated in \citet{roelfsema2012},
since the percentage differences between both works range only from 2$\%$ to 0$\%$ with increasing frequency. From the measured beam patterns, power 
distributions, and encircled energy fraction (EEF) in \citet{mueller2014} there is a substantial fraction of the power in the side lobes 
that varies with frequency. This is shown by the slow convergence of the EEF at large radii (> 20$\times$$\Theta_{FWHM}$). 
In a recent publication, \citet{volker2013} used the  antenna temperature scale for \CII\textrm{ }observations 
arguing for a large contribution from the side lobes to their extended measured emission. For the NANTEN2/SMART telescope, the error beam  at 461.1 GHz 
is expected to be within 7 to 10 times the size of the  main lobe  \citep[Internal NANTEN2 Report]{simon2007}, i.e. at radii less than 3'. The  error beam 
is expected to contain around 20\% of the detected power (U. Graf, private communication). Given the extended emission of the source and the location of 
the error beams for both telescopes, we consider that the most representative temperature scale of the true convolved antenna temperatures within 
the Sgr A region is the \emph{Forward Beam Antenna Temperature $T_{A}^{*}$}, where the corresponding forward efficiency 
$F_{eff}=\iint\limits_{2\pi} P(\Omega)\,d\Omega / \iint\limits_{4\pi} P(\Omega)\,d\Omega$ \citep{kramer1997}, 
with $P(\Omega)$ the antenna pattern of the telescope, has been applied. We apply this temperature scale to all data sets used in the present work. 
Given the contribution of the error beam pickup, the $T_{A}^{*}$ scale represents a lower limit to the true convolved antenna temperature. When the 
emission is more confined inside the respective error beam, the \emph{Main Beam Antenna Temperature} scale, with the corresponding main beam (MB) 
efficiency $B_{eff}=\iint\limits_{MB} P(\Omega)\,d\Omega / \iint\limits_{4\pi} P(\Omega)\,d\Omega$ \citep{kramer1997}, would imply a 
negligible contribution of the error beam pickup to the measured antenna temperatures, increasing the line strength by $\frac{F_{eff}}{B_{eff}}$.\\

\subsubsection{Cross-calibration}\label{crosscal}

The \CILINEAI\textrm{ }line was used to cross-check the antenna temperatures measured with both telescopes. A small area around the peak intensity,
shown by the Herschel-HIFI observations, was selected and re-observed with the NANTEN2/SMART telescope. The spatial resolution of the map is 46'' 
(smoothed to the Herschel-HIFI resolution) of a spacing of 23''$\times$23''. The spectral resolution of the data is 1 \KMS\textrm{ }and the rms noise 
temperature ranges from 0.044 K up to 0.071 K. The number of spectra used for the cross-calibration around the emission peak is 20. 
In order to compare the antenna temperatures measured with the two telescopes at the same positions on the sky, we explore the behaviour of the 
integrated intensity ratio with integration velocity interval. To minimize the influence of the noise on our results, we select 
only the emission that is above a 10$\sigma$ detection level, where $\sigma =$ 0.071 K is the maximum rms noise in the observed spectra. 
This threshold defines a velocity range $\Delta V$ from $+$23 \KMS\textrm{ }to $+$83 \KMS\textrm{ }common for all spectra.\\

For each measured spectrum we divide $\Delta V$ into velocity intervals ($\Delta V_{i}$) in order to calculate the total integrated intensity in each of them. 
The following values were used for $\Delta V_{i} $: 1, 2, 3, 4, 5, 6, 10, 15, 20, 30, and 60 \KMS. The velocity range $\Delta V$ is divided into a given 
$\Delta V_{i}$ value, yielding $n = \frac{\Delta V}{\Delta V_{i}}$ sub-intervals. With these definitions, the integrated intensity ratio $R(\Delta V_{i})_{j}$ in 
the $j$ sub-interval for the given $\Delta V_{i}$ value is calculated from the Herschel-HIFI $I(\Delta V_{i})_{j}^{H}$ and the NANTEN2/SMART $I(\Delta V_{i})_{j}^{N}$ 
integrated intensities as $R(\Delta V_{i})_{j} = I(\Delta V_{i})_{j}^{N}/I(\Delta V_{i})_{j}^{H}$. Then, the average integrated intensity ratio $R(\Delta V_{i})$ for 
the given $\Delta V_{i}$ interval is calculated as $R(\Delta V_{i}) = \frac{1}{n}\sum_{j=1}^n R(\Delta V_{i})_{j}$ for each pair of spectra observed at position 
$k$.\\

\begin{figure}
\centering
\includegraphics[angle=90, width=\hsize]{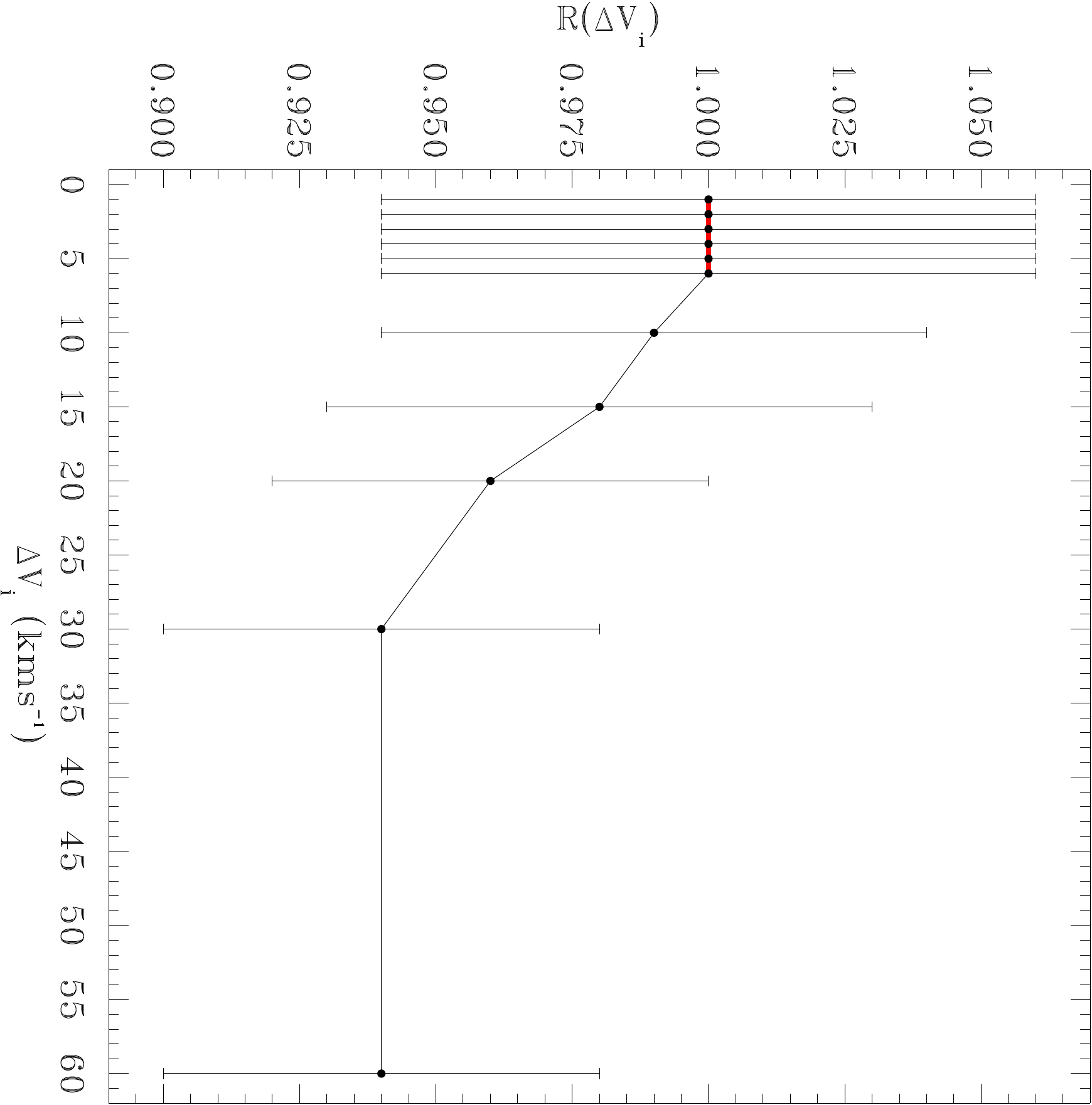}
\caption{Averaged integrated intensity ratio R$(\Delta V_{i})$ of the \CILINEAI\textrm{ }emission measured with the Herschel-HIFI and NANTEN2/SMART telescopes as a 
function of velocity integration interval $\Delta V_{i}$ as explained in the text. The intensity ratios are on average $\sim$ 2\% lower (within the error uncertainties) 
than unity over the whole range of $\Delta V_{i}$ values, showing that the measured NANTEN2/SMART antenna temperatures are systematically lower than those 
measured with the Herschel-HIFI telescope by the same amount. The red line delimits the $\Delta V_{i}$ range where the R$(\Delta V_{i})$ values are closest to unity.}
\label{fig_app:cross_cal}
\end{figure}

We combine the integrated intensity ratios R$(\Delta V_{i})_{k}$ and $\Delta V_{i}$ information of all observed spectra. First, we calculate the averaged 
integrated intensity ratio for all spectra pairs at $\Delta V_{i}$ as $R(\Delta V_{i}) = \frac{1}{N}\sum_{k=1}^N R(\Delta V_{i})_{k}$, with $N = 20$. From this, 
we obtain one $R(\Delta V_{i})$ value for a given $\Delta V_{i}$ interval. The results are shown in Figure \ref{fig_app:cross_cal}. The error bars correspond 
to the root-mean-square variation of each point and vary from 6\% of R$(\Delta V_{i})$) at $\Delta V_{i} = $ 1 \KMS\textrm{ }to 4\% at $\Delta V_{i} =$ 60 \KMS.\\ 

From the figure, R$(\Delta V_{i})$ is constant at 1.0 (red line) for $\Delta V_{i} \leq$ 6 \KMS. For larger $\Delta V_{i}$ values there is a small decrease to 0.94,
which could be attributed to differences in the error beam extent between the two telescopes, causing the measured antenna temperatures to be the result of the 
emission measured over slightly different spatial scales. The measured NANTEN2/SMART integrated antenna temperatures are consistently lower ($\sim$ 2\% on average) 
than the measured quantities in the Herschel-HIFI observations. For comparison, assuming a calibration error of 20\% in the absolute antenna temperature measured 
with both telescopes and an integrated intensity ratio of 0.98, the corresponding error is $\sim$ 30\%, much larger than the 2\% systematic underestimation of the 
Herschel-HIFI measured antenna temperatures by the NANTEN2/SMART measurements. Therefore, we are confident that both telescopes measure basically the same antenna 
temperatures on the $T_{A}^{*}$ scale;  given the large spatial extension of the emission,  we consider this a fair representation of the true convolved antenna 
temperature of the source.\\

\subsubsection{ASTE [CI]1-0 CMZ survey}

Recently, \citet{tanaka2011} carried out a 34'' angular resolution \CILINEAI\textrm{ }Survey of the CMZ using the 10 m Atacama Sub-millimeter Telescope Experiment (ASTE) 
telescope. At a spatial resolution of 46'' and spectral resolution of 2 \KMS, the \CILINEAI\textrm{ }peak intensity in their data set is 10.9 $\pm$ 0.2 K, at 
(l, b) $=$ ($-$0\DEG.02, $-$0\DEG.07), at LSR velocity $+$53 \KMS\textrm{ }(K. Tanaka, private communication) in the main beam antenna temperature scale 
($B_{eff} =$ 0.50). From 800 GHz measurements, the forward efficiency of the ASTE is considered to be $F_{eff} \sim$ 1.0 and the error beam pickup was estimated to be 
10\% of the power peak detected by the antenna. The absolute calibration error is estimated to be around 8\%. In the $T_{A}^{*}$ antenna temperature scale, their peak 
intensity reduces to 5.45 $\pm$ 0.10 K.\\

The peak intensity of Herschel-HIFI \CILINEAI\textrm{ }observations with 2 \KMS\textrm{ }spectral resolution and 46'' spatial resolution is 8.13 $\pm$ 0.05 K in 
$T_{A}^{*}$ temperature scale, at position (l,b) $=$ ($-$0\DEG.017, $-$0\DEG.071) in Galactic coordinates and at LSR velocity $+$53 \KMS. In the case of the 
\CILINEAI\textrm{ }NANTEN2/SMART observations used for the cross-calibration with the Herschel-HIFI data, at the same position and LSR velocity, we measured 
6.09 $\pm$ 0.06 K in $T_{A}^{*}$ temperature scale. This value is consistent with the Herschel-HIFI value within error uncertainties when considering the absolute 
calibration error. When considering the absolute calibration errors in the Herschel and ASTE data sets, the measured intensity peaks in $T_{A}^{*}$ scale are not 
consistent within error uncertainties. Nonetheless, we noticed that a lower forward efficiency in the case of the ASTE telescope would bring their measured intensity 
peak closer to the Herschel-HIFI value, and since $F_{eff}$ was derived only from 800 GHz measurements, this could indeed be  the case for the 492 GHz observations.\\ 

\subsection{Observations summary}\label{summary_obs}

A summary of the main parameters describing the scientific version data sets of all observed lines is presented in Table \ref{tab:data_summary}. 
The central position of all Herschel-HIFI observed maps is $\alpha(J2000) =$ 17$^{h}$\textrm{ }45'\textrm{ }39.9'', 
$\delta(J2000) = -$29\DEG\textrm{ }00'\textrm{ }28.1'' in equatorial coordinates, while the NANTEN2/SMART observations are centred at 
$l =$ 0\DEG.000, $b =$ 0\DEG.000 in Galactic coordinates for historical reasons (observations started as a  larger CMZ survey). The columns in 
Table \ref{tab:data_summary} counted from left to right contain \hypertarget{tel}{telescope name}; \hypertarget{linea}{observed line name}; 
\hypertarget{freq}{transition frequency ($\nu_{obs}$)}; \hypertarget{beam}{telescope beam size at the observed frequency ($\Theta_{FWHM}$)} taken 
from \citet{roelfsema2012} in the case of Herschel-HIFI observations; \hypertarget{beamfinal}{final spatial resolution of the data set ($\Theta$)}; 
\hypertarget{feff}{forward efficiency of the telescope ($F_{eff}$)}; \hypertarget{beff}{main beam efficiency of the telescope ($B_{eff}$)} from 
\citet{mueller2014} in the case of Herschel-HIFI observations; \hypertarget{rmsnoise}{typical noise in the data set (T$_{A,rms}^{*}$)};
\hypertarget{lsrvel}{original LSR velocity resolution ($\Delta V_{res}$)}; \hypertarget{resres}{resampled spectral resolution ($\Delta V_{f}$)}; 
\hypertarget{radvel}{LSR velocity coverage of the observations ($\Delta V_{range}$)}; and \hypertarget{energia}{energy of the upper level above 
the ground state for the observed line ($E_{u}$)}. In the next section, the morphology of the emission in all lines as a function of LSR 
velocity is described and discussed. \\

\begin{sidewaystable*}
\centering
\caption[Summary of Observations]{Summary of data sets after data reduction process.}
\begin{tabular}{llcccccccccc}
\hline
\hyperlink{tel}{Telescope}            &  
\hyperlink{linea}{Line}               &  
\hyperlink{freq}{$\nu_{obs}$}          & 
\hyperlink{beam}{$\Theta_{FWHM}$}      & 
\hyperlink{beamfinal}{$\Theta$}       & 
\hyperlink{feff}{$F_{eff}$}            & 
\hyperlink{beff}{$B_{eff}$}            & 
\hyperlink{rmsnoise}{T$_{A,rms}^{*}$}   & 
\hyperlink{lsrvel}{$\Delta V_{res}$}   & 
\hyperlink{resres}{$\Delta V_{f}$}     & 
\hyperlink{radvel}{$\Delta V_{range}$}  & 
\hyperlink{energia}{$E_{u}$}           \\ 
           &        & (GHz)          & ('')            &  ('')     &          &          & (K)   & (\KMS)         &  (\KMS)         &      (\KMS)       & (K)  \\
\hline
NANTEN2/SMART & \COLINEAIl & 461.0   & 37.4            &  40.0     & 0.86     & 0.50     & 0.14 &  0.677\tablefootmark{a}/0.050\tablefootmark{b}   &  1.0  & $-$200 to $+$200  & 55 \\
Herschel-HIFI & \CILINEAIl & 492.2   & 43.1            &  46.0     & 0.96     & 0.62     & 0.06 &  0.305         &  1.0            & $-$200 to $+$200  & 25 \\
Herschel-HIFI & \CILINEAIIl& 809.3   & 26.2            &  28.0     & 0.96     & 0.63     & 0.13 &  0.185         &  1.0            & $-$200 to $+$200  & 60 \\
Herschel-HIFI & \NIIl      & 1461.1  & 14.5            &  46.0     & 0.96     & 0.58     & 0.28 &  0.103         &  1.0            & $-$200 to $+$94   & 70 \\
Herschel-HIFI & \CIIl      & 1900.5  & 11.2            &  46.0     & 0.96     & 0.59     & 0.42 &  0.079         &  1.0            & $-$200 to $+$144  & 90 \\
\hline
\end{tabular}\label{tab:data_summary}
\tablefoot{
\tablefoottext{a}{For AOS backends}
\tablefoottext{b}{For XFFTS backends.}
}
\end{sidewaystable*}


\section{Emission morphology}\label{sources}

In the following, we discuss the spatial distribution of the emission in all lines, and across the $\pm$ 200 \KMS\textrm{ }LSR velocity range
in our data. The data are presented in channel maps of integrated intensity in units of K \KMS\textrm{ }($I = \int T_{A}^{*}(v)\textrm{ }dv$). For each map, the 
central LSR velocity is given. The spatial resolution in each data set is shown as a filled black circle in each map. For all observations,
the Right Ascension (R.A.) and Declination (DEC.) offsets, and Galactic longitude (l) and Galactic latitude (b) offsets are with respect to the central
position mentioned in Section \ref{summary_obs}. The channel maps can be found in \citet{garcia2015}. Given the large spatial and angular extent of our data sets, 
extra channel or integrated intensity maps are displayed when necessary for a better overview of the discussion of particular astronomical sources. In those figures, 
the NANTEN/SMART data have been rotated to the equatorial (J2000) coordinate system for better comparison with the Herschel-HIFI observations.\\

\begin{figure}[!h]
\centering
\includegraphics[angle=-90, width=\hsize]{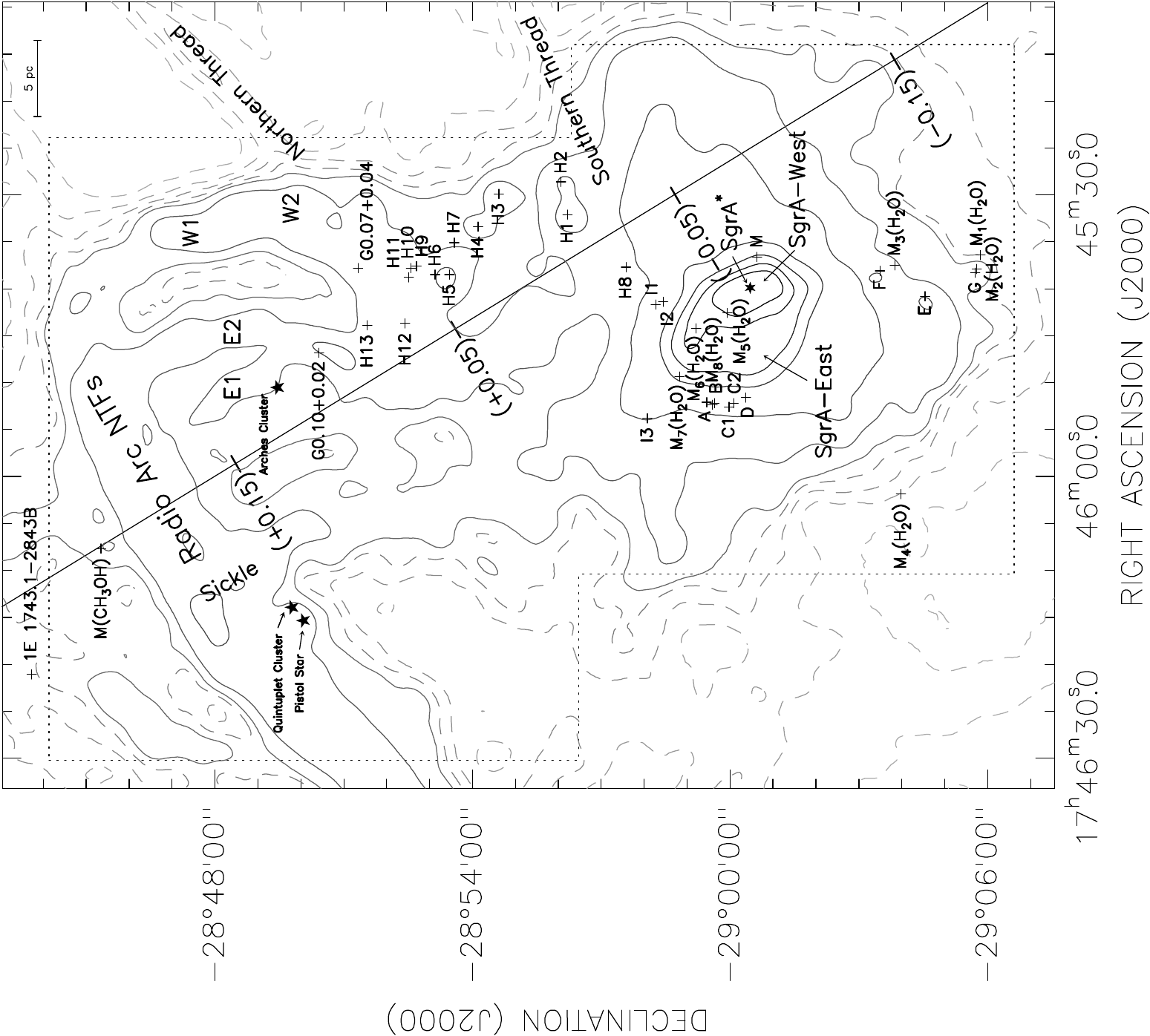}
\caption{20 cm continuum observations at 30'' spatial resolution from \citet{yusef1987a}. Overplotted on the figure are several discrete sources
found in the literature (see Section \ref{sources}). Following \citet{lang2010}, contours levels are at 10, 15, 20, 25, 30, 35, 50, 100, 300, 500, 700, 
and 900 times 8.5 mJy beam$^{-1}$. The solid straight line shows the position of the Galactic plane at b $= 0$\DEG, with marks denoting a few 
Galactic longitude values for orientation. The dotted line shows the area covered by the Herschel-HIFI observations.}
\label{continuum_yusef}
\end{figure}

\begin{figure}[!h]
\centering
\includegraphics[angle=-90, width=\hsize]{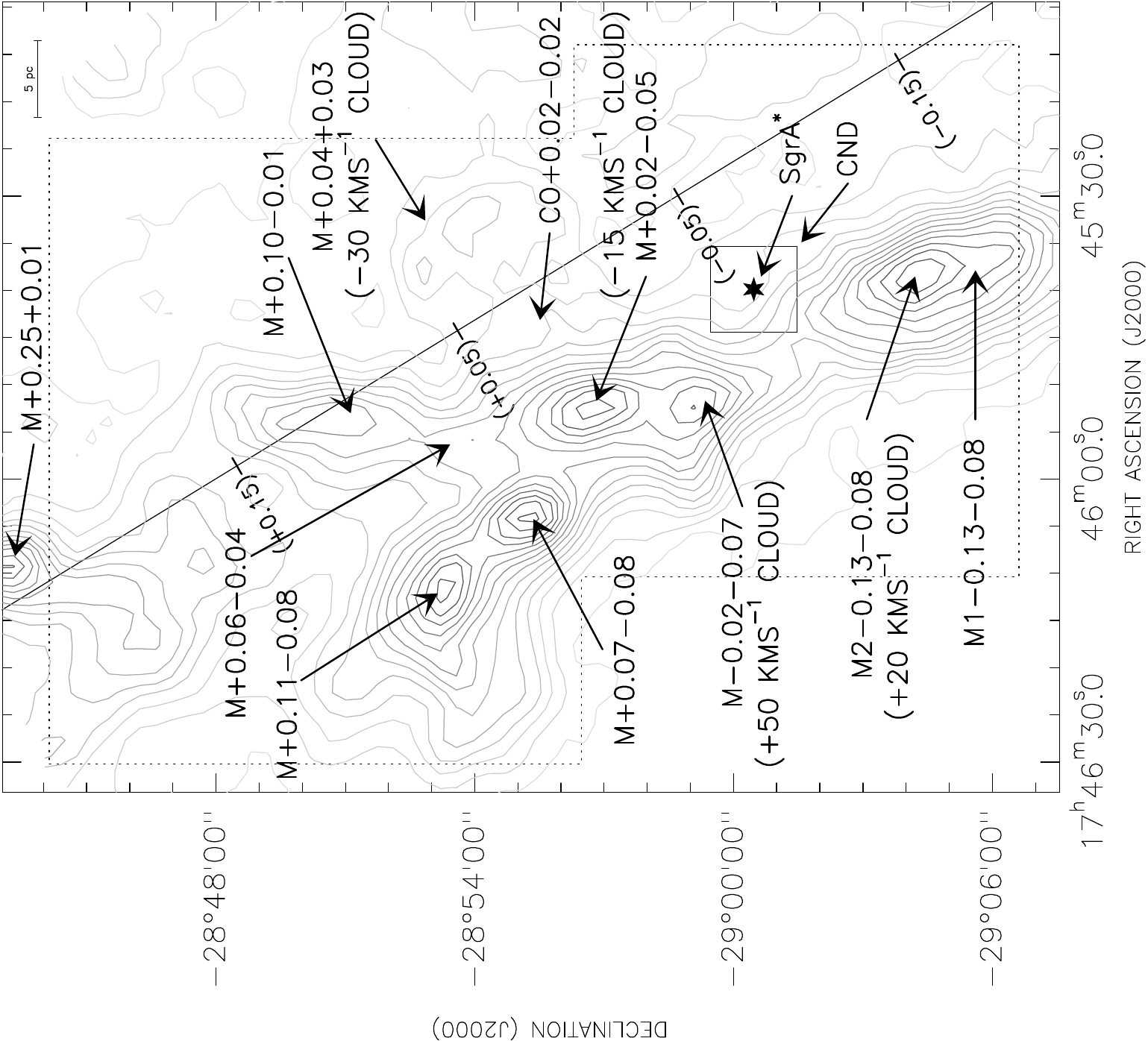}
\caption{\DIAZENYLIUM\textrm{ }(diazenylium) integrated intensity map between $-$75 and +110 \KMS\textrm{ } of the Mopra observations by \citet{jones2012}. The spatial 
resolution of the map is 44'' with 3.6 \KMS\textrm{ }spectral resolution. The first contour is located at $\sim$ 1 K \KMS\textrm{ }, equivalent to 5$\times \sigma$ with
$\sigma$ the typical noise of the map. The contour increment is 5 K \KMS. The solid straight line shows the position of the Galactic plane at b $= 0$\DEG; 
the marks denote a few Galactic longitude values for orientation. The area is covered  by the Herschel-HIFI observations.}
\label{diazenylium_jones}
\end{figure}

As described in Section \ref{Intro}, the Sgr A Complex contains an enormous variety of astronomical sources. In order 
to give a better overview of the region and the sources within it, we have compiled  a number of relevant sources from the literature 
for the discussion on the morphology of the region in our sub-mm/FIR observations \citep{guesten1981,serabyn1987,yusef1987a,
yusef1987b,zhao1993,cotera1996,timmermann1996,lang1997,lang1999a,lang1999b,lang2001,lang2002,porquet2003,oka2008,caswell2010,lang2010,caswell2011,requena2012}. 
Figure \ref{continuum_yusef} shows the 20 cm continuum emission observed by \citet{yusef1987a} delineating large-scale structures and point sources. 
The figure includes the position of the massive black hole Sgr A$^{\star}$ (very close to which the massive Nuclear cluster is found), 
the A-I and M \HII\textrm{ }regions close to the GC; the Sgr A-East and Sgr A-West regions; the Radio Arc and other non-thermal structures such as the Northern and Southern 
Threads; the Thermal Arched Filaments E1, E2, W1, W2; the ``banana'' G0.10$+$0.02 and G0.07$+$0.04 structures; the positions of the massive Quintuplet and Arches 
clusters, and of the Pistol star (G0.15$-$0.05); the extended Sickle \HII\textrm{ }region, several \HII\textrm{ }regions (H$+$number) forming the 
H Region; water and methanol masers; and the X-ray source 1E 1743.1$-$2843B. The dotted-line in the map shows the area covered by the Herschel-HIFI 
observations while the solid straight line indicates the Galactic plane at b $=$ 0\DEG. In order to have an overview of the molecular material in the Sgr A Complex, 
Figure \ref{diazenylium_jones} shows the integrated intensity map (between $-$75 \KMS\textrm{ }and $+$110 \KMS) of \DIAZENYLIUM\textrm{ }(diazenylium) of the 
Mopra observations by \citet{jones2012}, tracing cold and dense gas. In this figure, the position of all the major molecular clouds (including the ``Brick'' M$+$0.25$+$0.01), 
the HVCC (CO$+$0.02$-$0.02), and the extent of the CND (black square), are shown. The diversity of large structures seen in Figures \ref{continuum_yusef} and 
\ref{diazenylium_jones} shows the need of multiwavelength comparison in order to understand the large-scale spatial and spectral distribution of the warm gas 
component traced in our data.\\

\subsection{Average spectrum of the observed sub-mm lines}

\begin{figure}
\centering
\includegraphics[angle=-90, width=\hsize]{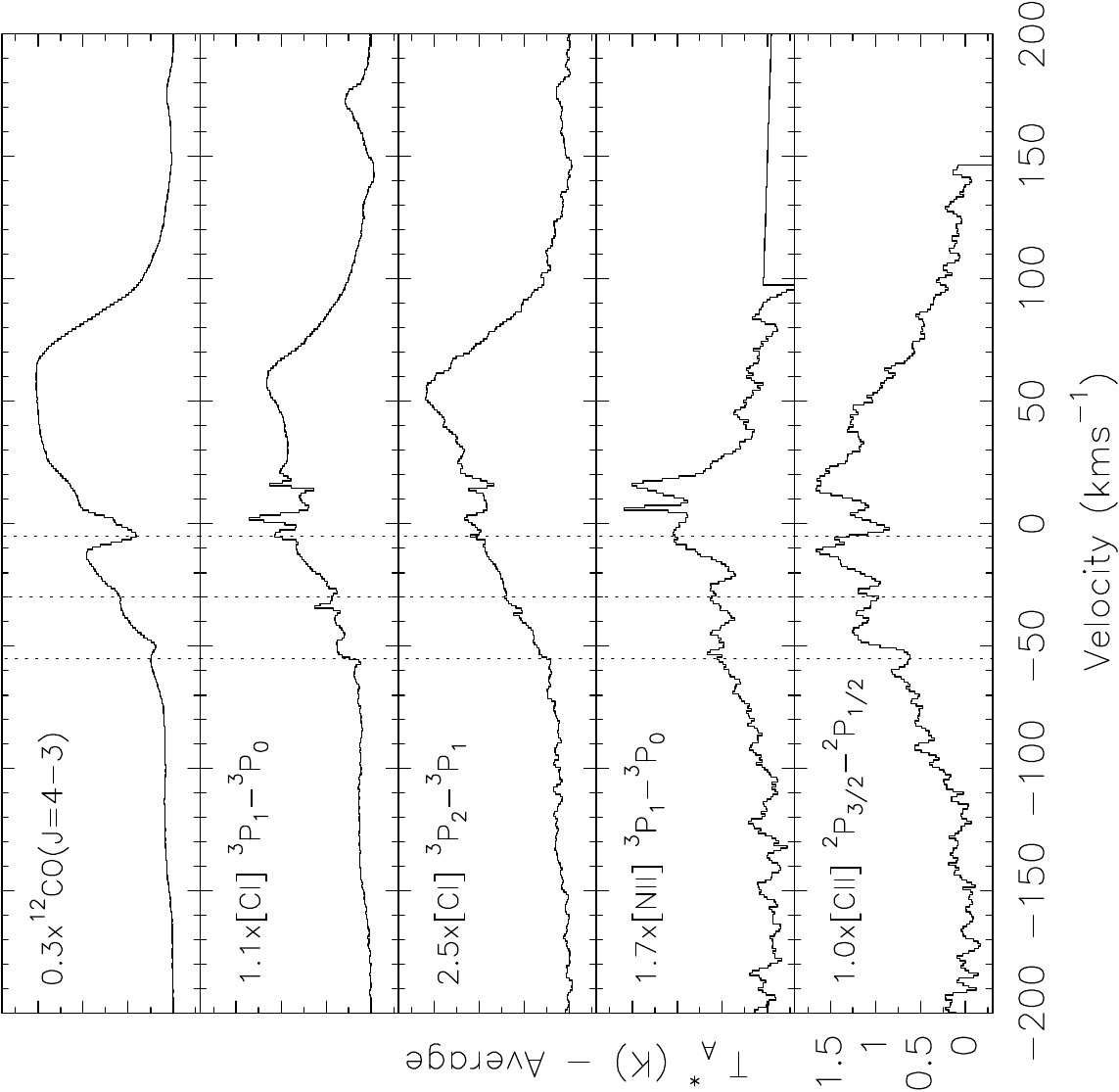}
\caption{Average spectra of the sub-mm/FIR lines observed in the present work. Each average spectrum is made out of 46'' spatial resolution data cubes 
and within the area covered by the Herschel-HIFI observations. Vertical dotted lines show the loci of the 3 kpc, 4.5 kpc, and local arm spiral features
at $\sim$ $-$55 \KMS, $\sim$ $-$30 \KMS, and $\sim$ $-$5 \KMS, respectively.}
\label{morpho:final_average_spectra}
\end{figure}

The average spectrum of the observed submillimetre lines is shown in Figure \ref{morpho:final_average_spectra}. The spectra were 
obtained by bringing all data sets to a common 46'' angular resolution and averaging all spectra within the area covered by 
the Herschel-HIFI observations. If one considers the average spectrum as a single-point observation, it would
be roughly equivalent to a spectrum taken with a spatial resolution of $\approx$ 50 pc. The bulk of the carbon monoxide and atomic carbon emission 
is concentrated in the major MCs towards positive LSR velocities, while the emission of the ionized material is more symmetric with respect to 0 \KMS. 
Given the low rms noise of the Herschel-HIFI and NANTEN2/SMART observations, even weak emission at very high 
LSR velocities (\VLSR\textrm{ }$>$ $+$150 \KMS) in the \CILINEAI\textrm{ }and \COLINEAI\textrm{ }lines is detected in the average spectrum. 
Absorption features along the l.o.s. associated with the loci of spiral arms are observed in the \COLINEAI\textrm{ }average spectrum: at $-$5 \KMS\textrm{ }the most 
prominent absorption feature associated with the local arm is seen, while other weaker absorption features at $-$30 \KMS\textrm{ }and $-$55 
\KMS\textrm{ }are associated with the 3 kpc and 4.5 kpc spiral arms \citep{oka1998,jones2012,dame2001,dame2008,garcia2014}.
The average \CII\textrm{ }spectrum also shows some correspondence with the absorption LSR velocity of the spiral arms along the l.o.s.,
while no clear absorption signatures are seen in the carbon average spectra. It is important to notice that the absorption due to
foreground spiral arms is expected to be narrow with $\Delta V \leq$ 5 \KMS\textrm{ }for the CO(1-0) transition \citep{oka1998}.
Therefore, broader absorption features in the warm gas could (at least partially) originate from in situ absorption. For instance,
the absorption feature at $-$5 \KMS\textrm{ }seen in the average \COLINEAI\textrm{ }spectrum has a line width $\Delta V \approx$ 10 \KMS.\\

\subsection{Molecular Cloud Complexes between $-$80 \KMS up to $+$90 \KMS}\label{molecular_clouds}

\begin{figure*}[!t]
\begin{minipage}{\hsize}
\centering
\includegraphics[angle=-90, width=\hsize]{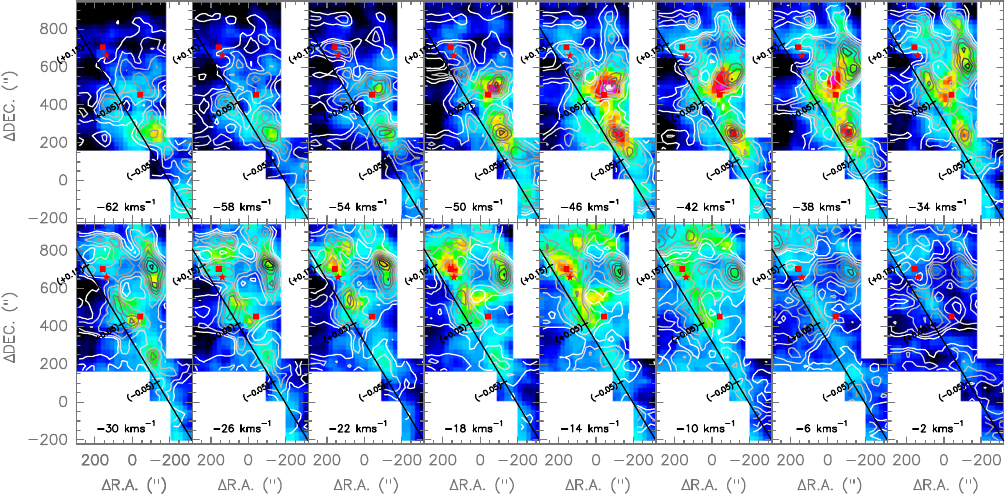}
\end{minipage}
\begin{minipage}{\hsize}
\centering
\includegraphics[angle=-90, width=\hsize]{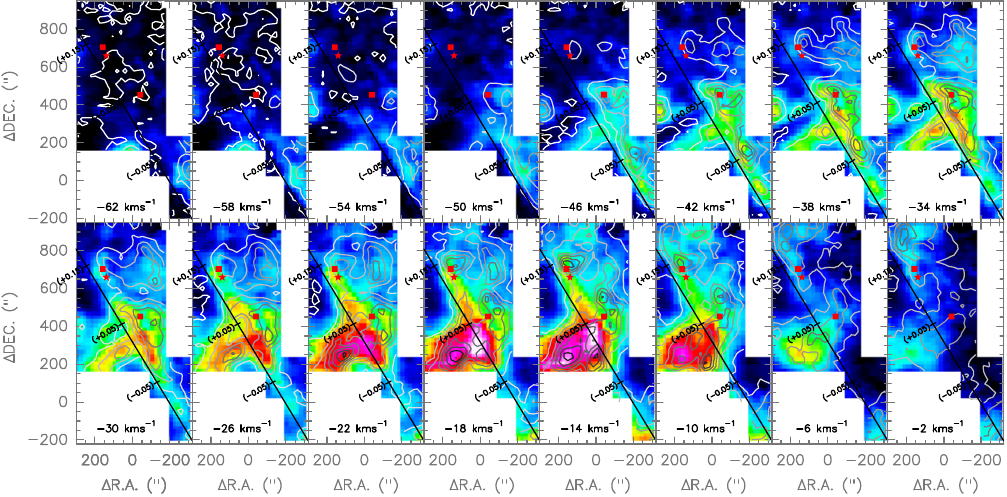}
\caption{Integrated intensity maps (5 \KMS\textrm{ }wide) of the \CII\textrm{ }(colour scale) and \NII\textrm{ }(contours) lines in the top panels and of the 
\COLINEAI\textrm{ }(colour scale) and \CILINEAI\textrm{ }(contours) lines in the bottom panels, tracing the molecular and atomic material towards the Arched 
Filaments E1, E2, W1, and W2 and the H Region as shown in Figure \ref{morpho:final_average_spectra}. The red star shows the location of the Arches Cluster, 
while red squares show the positions of two CS(2-1) peaks shown in the work by \citet{serabyn1987}, tracing high density material. The solid straight 
line shows the position of the Galactic plane at b $=$ 0\DEG, with marks denoting a few Galactic longitude values for orientation. The central LSR velocity 
of the maps is shown in each panel.}
\label{arches_CO_CI}
\end{minipage}
\end{figure*}

Most of the molecular clouds within the Sgr A Complex are located between $-$80 \KMS\textrm{ }and $+$90 \KMS. From negative to positive LSR velocities,
the first  feature is M$+$0.04$+$0.03 ($-$30 \KMS\textrm{ }cloud), which contains the gas associated with the Arched Filaments 
\citep{serabyn1987,zhao1993} ranging from $\sim$ $-$70 \KMS\textrm{ }up to 0 \KMS\ and covering the upper half of the Herschel-HIFI maps. Figure 
\ref{arches_CO_CI} (upper panels) shows 5 \KMS\textrm{ }integrated intensity maps of the \CII\textrm{ }(colour scale) and \NII\textrm{ }(contours) emission
with the corresponding central LSR velocity of the maps shown in the bottom of each panel. The red squares depict the positions of two peaks in CS(2-1) 
emission detected by \citet{serabyn1987}, tracing high density material, while the red star shows the position of the massive Arches Cluster. 
The emission in both lines is very widespread and closely follows the 20 cm continuum emission shown in Figure \ref{continuum_yusef}, moving from the H Region, 
with a local intensity peak in both lines at the position of the H1-H2 sources, to the Arched Filaments, going through G0.07$+$0.04, and moving from the W1/W2 
filaments to the E1/E2 filaments and the ``banana'' as the LSR velocity approaches 0 \KMS. From the maps, a strong spatial correlation between the 
\CII\textrm{ }and \NII\textrm{ }lines is observed, indicating that there is a significant contribution to the observed \CII\textrm{ }emission from the  
\HII\textrm{ }regions and not only from PDRs \citep{abel2005}. The case of  G0.07$+$0.04 is particularly interesting since it is thought that  
at  this position the gas is interacting with the Northern Thread \citep{lang1999b}. In a preliminary analysis, the central LSR velocity of 
the H92$\alpha$ average spectrum in this region shown by \citet{lang2001} is shifted by around $\sim$ 30 \KMS\textrm{ }towards more negative velocities 
with respect to the bulk of carbon and carbon monoxide emission along the same l.o.s. Such a large velocity shift is not seen at other positions of the 
Arched Filaments, and might indicate that the ionized material is drifting away from G0.07$+$0.04 as a result of the interaction with the local magnet field.

\begin{figure}[!h]
\begin{minipage}{\hsize}
\centering
\includegraphics[angle=-90, width=\hsize]{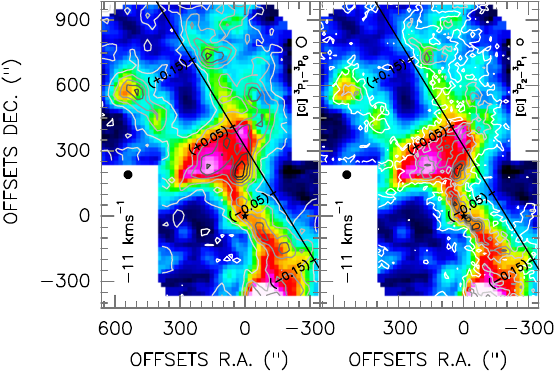}
\end{minipage}
\begin{minipage}{\hsize}
\centering
\includegraphics[angle=-90, width=\hsize]{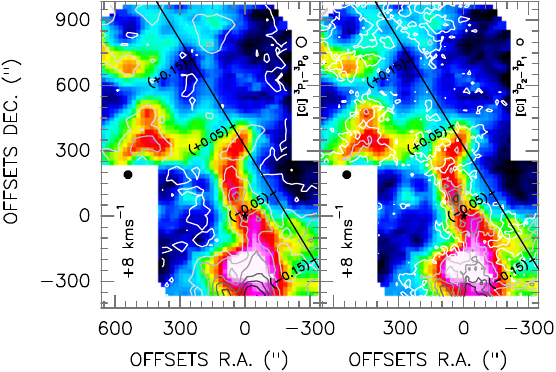}
\caption{Channel maps at $-$11 \KMS\textrm{ }(top panels) and $+$8 \KMS\textrm{ }(bottom panels) of the \COLINEAI\textrm{ }(colour scale) emission, while  
the \CILINEAI\textrm{ }(left panels) and \CILINEAII\textrm{ }(right panels) lines are shown as contours. Open circles show the spatial resolution of the 
carbon lines, while the black filled circles show the spatial resolution of the carbon monoxide line. The $\star$ symbol shows the position of Sgr A$^{\star}$.
The colour scale is the same in all panels. The solid straight line shows the position of the Galactic plane at b $=$ 0\DEG;  the marks denote a few 
Galactic longitude values for orientation.}
\label{streamer_-11_+8_kms}
\end{minipage}
\end{figure}

The peak \CII\textrm{ }intensity ($\sim$ 9.2 $\pm$ 0.4 K) occurs within the Arched Filaments at $\Delta \alpha(J2000) =$ $-$82.2'', $\Delta \delta(J2000) =$ $+$490.4'', 
and $V_{LSR} =$ $-$45 \KMS\textrm{ }, while the peak intensity of the \NII\textrm{ }line ($\sim$ 2.7 $\pm$ 0.4 K) occurs at  $\Delta \alpha(J2000) =$ $-$107.2'', 
$\Delta \delta(J2000) =$ $+$720.3'', and $V_{LSR} =$ $-$26 \KMS. Figure \ref{arches_CO_CI} (bottom panels) shows the emission distribution of the 
\COLINEAI\textrm{ } and \CILINEAI\textrm{ }. The behaviour of the \CILINEAII\textrm{ }line is very similar to the \CILINEAI\textrm{ }emission 
(as expected), so we focus the discussion on the lower frequency line. At large negative LSR velocities, there is a gas streamer connecting the H region with 
the lower part of the Arched Filaments, where one of the CS(2-1) peaks is located (see Section \ref{streamers}). The bulk of the emission detected from the 
\COLINEAI\textrm{ }and \CILINEAI\textrm{ }lines is displaced with respect to the position of the filaments and with respect to the CS(2-1) emission 
peaks  detected by \citet{serabyn1987}, and it is weak in comparison to the rest of the emission in the data cubes. The position of the most 
intense emission in the \COLINEAI\textrm{ }line is at the centre of the maps, as seen for example at $-$18 \KMS, between M$+$0.04$+$0.03 ($-$30 \KMS\textrm{ }cloud) 
and M$+$0.02$-$0.05 ($-$15 \KMS\textrm{ }cloud), while local maxima of the carbon emission coincide with the position of these two clouds.
In the top panels of Figure \ref{streamer_-11_+8_kms}, at $-$11 \KMS, a lane of emission (both in carbon and carbon monoxide) is seen going from the  position 
of M$+$0.02$-$0.05 through the Galactic Centre and connecting with another cloud south of the CND, which is probably part of the $+$20 \KMS\textrm{ }cloud. \citet{serabyn1987} suggested that the lack of ionized material in the region around M$+$0.02$-$0.05, the so-called {molecular bridge}, could be 
the result of the tidal disruption of material in infall towards the CND. This seems to be supported by the morphology of the warm gas in our observations
(see Section \ref{streamers}).\\

From 0 \KMS\textrm{ }to $+$90 \KMS, there is a large number of molecular clouds outlined in different temperature and density tracers at lower frequencies. 
We refer to them following the nomenclature used by \citet{guesten1981} who identified their \AMONIA\textrm{ }peak positions, as we did in Figure \ref{diazenylium_jones}.
Given the molecular nature of the MCs, these are brighter in the \COLINEAI\textrm{ }emission than in any other line in our data. From 0 \KMS\textrm{ }to 
$+$20 \KMS, the brightest feature is the M2$-$0.13$-$0.08 ($+$20 \KMS\textrm{ }cloud) seen at $\Delta$l $<$ 0''. The emission lane that crosses the GC from the 
{molecular bridge} to connect with the $+$20 \KMS\textrm{ }cloud is still clearly visible at $+$8 \KMS, as seen in the bottom panels of Figure 
\ref{streamer_-11_+8_kms}. These streamers are traced almost uninterruptedly by the \CILINEAI\textrm{ }and the \CILINEAII\textrm{ }emission, since no  
absorption features are present in these lines within the $-$11 \KMS\textrm{ }to $+$8 \KMS\textrm{ }LSR velocity range. From the morphological point of view,  
this suggests that the molecular bridge and the $+$20 \KMS\textrm{ }cloud are part of the same structure that is being tidally disrupted in the gravitational field of 
the Nuclear Cluster, which dominates the gravitational potential for radii 2 - 30 pc \citep{longmore2013}. In this case, the emission gap between the bridge and the 
$+$20 \KMS\textrm{ }cloud in the \COLINEAI\textrm{ }emission is, at least partially, due the strong absorption of the local arm along the 
l.o.s.\\

\begin{figure}[!h]
\centering
\includegraphics[angle=-90, width=\hsize]{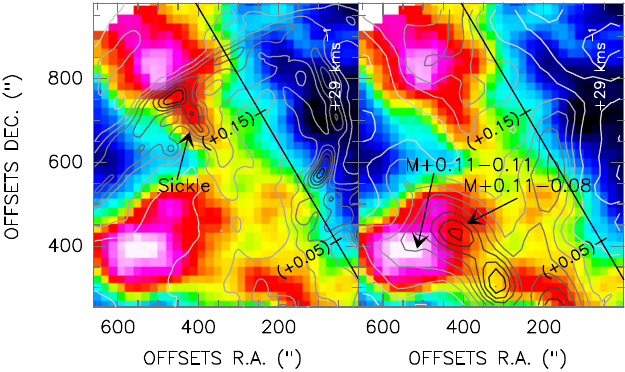}
\caption{Integrated intensity map (3 \KMS\textrm{ }wide) of the \COLINEAI\textrm{ }emission (colour scales) centred at $+$29 \KMS\textrm{ }with superimposed 
contours of the 20 cm continuum emission measured by \citet{yusef1987a} showing the Sickle \HII\textrm{ }region (left panel), and \DIAZENYLIUM\textrm{ }contours 
of the integrated intensity map between $-$75 and $+$110 \KMS\textrm{ } from the Mopra observations by \citet{jones2012} showing the emission 
distribution of the M$+$0.11$-$0.08 cloud and the position of the  M$+$0.11$-$0.11 cloud from \citet{tsuboi1997} (right panel). The solid straight 
line in both panels shows the position of the Galactic plane at b $=$ 0\DEG, with marks denoting a few Galactic longitude values for orientation. The 
\COLINEAI\textrm{ }emission related to the Sickle is located at the edge of the \HII\textrm{ }region while the \COLINEAI\textrm{ }emission related to the M$+$0.11$-$0.11 
cloud is shifted at this LSR velocity with respect to the peak emission of the cold gas traced by diazenylium.}
\label{sickle}
\end{figure}

For $\Delta$l $>$ 0'', two \COLINEAI\textrm{ }emission features appear at $\Delta$l $=$ $+$400'', $\Delta$b $=$ $-$400'', and $\Delta$l $=$ $+$750'', 
$\Delta$b $=$ $-$150''. The former surrounds the M$+$0.11$-$0.08  cloud as seen in \DIAZENYLIUM\textrm{ }emission from the Mopra observations \citep{jones2012},
and it is associated with the M$+$0.11$-$0.11 dense cloud observed in CS(1-0) and CS(2-1) by \citet{tsuboi1997}, 
while the latter is slightly shifted with respect to a local \DIAZENYLIUM\textrm{ }peak, not covered in the \citet{guesten1981} observations, and coincides 
spatially with the edge of the Sickle \HII\textrm{ }region. Both features anti-correlate with the 20 cm continuum emission measured by \citet{yusef1987a}. 
This is illustrated by the 3 \KMS\textrm{ }wide \COLINEAI\textrm{ }integrated intensity map shown in Figure \ref{sickle} at central LSR velocity $+$29 \KMS. 
The contours in the figure show the Sickle region depicted in the continuum emission (left panel), the M$+$0.11$-$0.08 cloud traced by the diazenylium 
line, and the position of the  M$+$0.11$-$0.11 cloud (right panel). Their \COLINEAI\textrm{ }emission extends up to $\sim$ $+$40 \KMS\textrm{ }where these structures are 
still recognizable. The atomic carbon line observations follow the same trend as the carbon monoxide observations in this LSR velocity range. There are \CII\textrm{ }and 
\NII\textrm{ }detections towards the Sickle \HII\textrm{ }region, while no clear detection is seen towards the M$+$0.11$-$0.08 cloud. In general, the 
\CII\textrm{ }emission is widespread, while the \NII\textrm{ }emission anti-correlates spatially at some spots with the lower frequency lines, as can be expected 
from gas adjacent to \HII\textrm{ }regions. An example of this is seen in the channel maps at LSR velocity $+$18 \KMS, where a local maximum in the 
\NII\textrm{ }emission is located just next to the \COLINEAI\textrm{ }emission of the $+$20 \KMS\textrm{ }cloud at $\Delta \alpha(J2000) \sim$ $-$150'', 
$\Delta \delta(J2000) \sim$ $-$250''.\\

\begin{figure}[!h]
\centering
\includegraphics[angle=-90, width=\hsize]{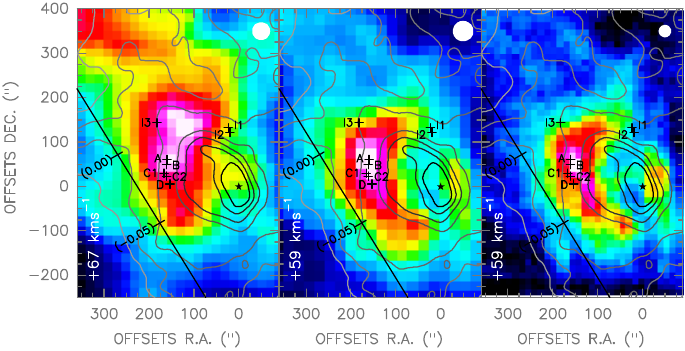}
\caption{Integrated intensity maps (3 \KMS\textrm{ }wide) of the \COLINEAI, \CILINEAI, and \CILINEAII\textrm{ }emission (colour scales) for 
the $+$50 \KMS\textrm{ }cloud. The filled white circles represent the spatial resolution of the data, and the central LSR velocity of the maps 
is given in each panel. The 20 cm continuum emission from the Sgr A-East region measured by \citet{yusef1987b} is shown as contours.
The solid straight line shows the position of the Galactic plane at b $=$ $-$0\DEG.10, with marks denoting a few 
Galactic longitude values for orientation. The crescent-shape in all three sub-mm lines at different LSR velocities surrounding the 
Sgr A-East region reveals the interaction of the $+$50 \KMS\textrm{ }cloud with the non-thermal source.}
\label{50_kms_cloud}
\end{figure}

Most of the neutral atomic carbon emission and molecular emission is found within the LSR velocity range $+$20 \KMS\textrm{ }up to $+$90 \KMS. The peak intensities
measured in our data sets are $T_{A}^{*} =$ 22.11 $\pm$ 0.09 K at $\Delta$l $=$ $-$64.9'', $\Delta$b $=$ $-$259.3'', and $V_{LSR} =$ $+$63 \KMS\textrm{ }for the 
\COLINEAI\textrm{ }line; $T_{A}^{*} =$ 8.14 $\pm$ 0.05 K at $\Delta \alpha(J2000) =$ $+$152.1'', $\Delta \delta(J2000) =$ $+$73.0'', and $V_{LSR} =$ $+$53 
\KMS\textrm{ }for the \CILINEAI\textrm{ }line; and $T_{A}^{*} =$ 6.98 $\pm$ 0.13  K at $\Delta \alpha(J2000) =$ $+$155.2'', $\Delta \delta(J2000) =$ $+$80.1'', 
and $V_{LSR} =$ $+$55 \KMS\textrm{ }for the \CILINEAII\textrm{ }line, all three associated in space and LSR velocity with M$-$0.02$-$0.07 ($+$50 
\KMS\textrm{ }cloud). For $\Delta$l $<$ $+$150'', when going from $+$20 \KMS\textrm{ }to $+$50 \KMS, the \COLINEAI\textrm{ }emission 
moves across Galactic longitude from the $+$20 \KMS\textrm{ }cloud ($\Delta$l $=$ $-$350'', $\Delta$b $=$ $-$250'') continuously to connect with the 
$+$50 \KMS\textrm{ }cloud at around $\Delta$l $=$ $-$350'', $\Delta$b $=$ $-$250''. This behaviour is also seen in the atomic carbon lines from which, 
at LSR velocities between $+$50 \KMS\textrm{ }and $+$60 \KMS, the $+$50 \KMS\textrm{ }cloud stands out as a crescent-shaped feature with its concave 
side oriented to the non-thermal shell source Sgr A-East, which is thought to be the remnant of a supernova explosion (or explosions) with energy
$\sim$ 4$\times$ 10$^{52}$ ergs \citep{yusef1987a,mezger1989,coil2000}. At $+$67 \KMS, a similar crescent-shaped structure in 
\COLINEAI\textrm{ }emission is seen. This is shown in Figure \ref{50_kms_cloud}, where the contours show the 20 cm  continuum emission and 
the position of Sgr A$^{\star}$ is represented by the $\star$ symbol. This indicates that the emission detected from the $+$50 \KMS\textrm{ }cloud is in close 
relationship with the energetic event that originated in the Sgr A-East region. The emission tracing the $+$50 \KMS\textrm{ }cloud extends up to $+$90 \KMS.\\

For $\Delta$l $>$ $+$150'', the \COLINEAI\textrm{ }emission between $+$20 \KMS\textrm{ }and $+$90 \KMS\textrm{ }traces the M$+$0.07$-$0.08, M$+$0.11$-$0.08, 
M$+$0.06$-$0.04, and M$+$0.10$-$0.01 molecular clouds and gas associated with the Sickle \HII\textrm{ }region, as mentioned before. 
The \COLINEAI\textrm{ }emission peak in this region is shifted with respect to the global \CILINEAI\textrm{ }intensity peak, around $+$57 \KMS, and 
is comparable in magnitude
to the one detected towards the $+$50 \KMS\textrm{ }cloud, having a similar morphology to the emission below $\Delta$l $=$ $+$150''
for LSR velocities between $+$56 \KMS\textrm{ }and $+$63 \KMS. This is not the case for the atomic carbon lines. The emission from the 
\CILINEAI\textrm{ }and \CILINEAII\textrm{ }lines above $\Delta \delta(J2000) =$ $+$250'' is much weaker towards these molecular clouds 
and shows a highly asymmetric spatial distribution. Given the symmetric distribution in the \COLINEAI\textrm{ }lines and the very 
asymmetric distribution of the \CILINEAI\textrm{ }and \CILINEAII\textrm{ }emission, it is likely that a different heating 
mechanism between the two regions is responsible for the gas excitation. \citet{minh2005} attributed the enhancement of the HCO$+$ emission in 
this region to the interaction of the gas with the shocks waves produced by the Sgr A-East supernova remnant. The \CII\textrm{ }emission, 
between $+$20 \KMS\textrm{ }and  $+$90 \KMS\textrm{ }is much weaker than at negative LSR velocities and in general follows the distribution 
of the molecular gas traced by \COLINEAI\textrm{ }below the Galactic plane.\\

\begin{figure}[!h]
\centering
\includegraphics[angle=-90, width=\hsize]{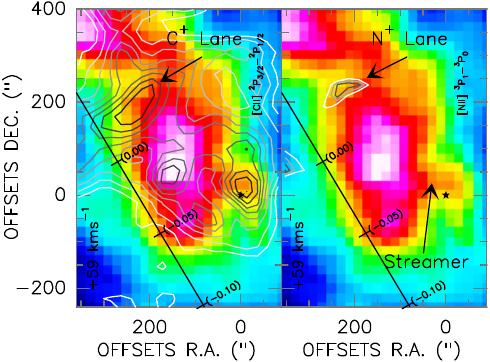}
\caption{Integrated intensity map (3 \KMS\textrm{ }wide) of the \COLINEAI\textrm{ }(colour scale) from the $+$50 \KMS\textrm{ }cloud centred at 
$+$58 \KMS\textrm{ }with \CII\textrm{ }(left) and \NII\textrm{ }(right) emission as contours overlaid on the maps. The first contour of the
high frequency lines is set at a 5$\times\sigma$ significance level, with $\sigma$ the typical noise of the map. The $\star$ symbol represents 
the position of Sgr A$^{\star}$. The solid straight line shows the position of the Galactic plane at b $=$ $-$0\DEG.10, with marks denoting a few 
Galactic longitude values for orientation. A prominent \CII\textrm{ }structure is found tangent to the convex side of the $+$50 \KMS\textrm{ }cloud with a 
small \NII\textrm{ }counterpart.}
\label{cii_emission_lane}
\end{figure}

Between $+$20 \KMS\textrm{ }and $+$70 \KMS, there is a prominent emission lane going from $\Delta \alpha(J2000) 
\sim$ $+$150'', $\Delta \delta(J2000) \sim$ $+$400'' to $\Delta \alpha(J2000) \sim$ $+$400'', $\Delta \delta(J2000) \sim$ $+$50''. 
The feature is tangent to the convex side of the $+$50 \KMS\textrm{ }cloud. Since the \NII\textrm{ }emission decreases significantly 
for LSR velocities above $+$20 \KMS, the same feature is barely detected in the \NII\textrm{ }line. This is shown in Figure 
\ref{cii_emission_lane}, where the \CII\textrm{ }and \NII\textrm{ }emission is overlaid on the \COLINEAI\textrm{ }emission
in colours. Again, the $\star$ symbol represents the position of Sgr A$^{\star}$. The composite 8.0, 4.5, and 3.6 $\mu m$ 
Spitzer/IRAC image of the Sgr A Complex in \citet{chambers2014} shows bright emission at the same position, indicating a 
lower opacity than towards the positions of the bulk molecular gas. The A-I \HII\textrm{ }regions associated with the $+$50 \KMS\textrm{ }cloud \citep{yusef1987a} 
are located towards the opposite side of the cloud and closer to the GC. We found no \HII\textrm{ }sources listed in the literature that correlate 
with the position of the emission. In general, the \NII\textrm{ }emission is more confined spatially than the \CII\textrm{ }emission, reflecting
their different origin;  \NII\textrm{ }emission is associated mainly with \HII\textrm{ }regions and \CII\textrm{ }emission tracing different 
stages of the ISM, including a contribution from \HII\textrm{ }regions.\\

It is interesting to note that most of the flux in the neutral atomic and molecular lines is located at Galactic latitudes below the Galactic plane (b $<$ 0\DEG), 
while the flux of the ionized material traced by the \NII\textrm{ }and \CII\textrm{ }peaks above the Galactic plane, running from $\Delta \alpha(J2000) =$ $-$350'', 
$\Delta \delta(J2000) =$ $-$300'' to $\Delta \alpha(J2000) =$ $+$450'', $\Delta \delta(J2000) =$ $+$1000'' in the Herschel-HIFI maps.
Given the offset of the dynamical centre of the Milky Way (Sgr A$^{\star}$) from the centre position of the Galactic coordinate system, the location
of the ionized and neutral material with respect to b $=$ 0\DEG\textrm{ }is not particularly meaningful, but the fact that both materials are found at different
positions in the Sgr A Complex could indicate a different evolutionary stage of the gas moving along a ring-like structure \citep{molinari2011}
whose passage through the closest point to the bottom of the gravitational potential could trigger the star formation process \citep{longmore2013}.\\

\subsection{Gas streamers}\label{streamers}

Gas streamers, first reported by \citet{ho1991} are seen at several locations in the GC. They are thought to be the result of 
material from MCs (such as the $+$20 \KMS\textrm{ }and $+$50 \KMS\textrm{ }clouds) trapped by the central gravitational potential. They are brighter 
in emission from warm gas than from cold gas as they are heated on their way towards the GC \citep{ho1991}. Using NH$_{3}$ observations, \citet{coil2000} reported 
the detection of the {southern streamer} stretching from the $+$20 \KMS\textrm{ }towards the CND but not reaching the GC. No detection of the 
northern part of the streamer is seen in their data, probably because of the low density of the gas. Nonetheless, the southern and northern 
parts of the streamer are clearly traced by \COLINEAI, \CILINEAI, and \CILINEAII\textrm{ }emission in our data, as  is shown in Figure \ref{streamer_-11_+8_kms}
at positive LSR velocities. In the \COLINEAI\textrm{ }line, the northern and southern parts of the streamer appear as a single continuous structure (at least in 
projection) going through the CND, and are remarkably similar to the streamer shown at negative velocities, also shown in Figure \ref{streamer_-11_+8_kms} (upper panels), 
suggesting that they could be physically related. The streamers would appear as separated entities only because of the accidental massive absorption feature at $-$5 \KMS.
If this were the case, it would imply that there is warm gas connecting the $-$15 \KMS\textrm{ }cloud with the $+$20 \KMS\textrm{ }cloud. In the same work, 
\citet{coil2000} showed that the $+$20 \KMS\textrm{ }and $+$50 \KMS\textrm{ }clouds are connected by a thin molecular ridge of gas (also referred to as the 
{eastern streamer}), not seen in dust maps, and suggested that both clouds may constitute part of the same large-scale structure of gas lying along the 
Galactic plane. Since NH$_{3}$ traces only the densest part  of the gas, they fail to detect intermediate densities of the warm gas traced by \COLINEAI. 
The eastern streamer is seen in \COLINEAI, \CILINEAI, and \CILINEAII\textrm{ }emission, in the velocity range $+$20 \KMS\textrm{ }to $+$50 \KMS, moving from 
the position of the $+$20 \KMS\textrm{ }cloud towards the location of the $+$50 \KMS\textrm{ }cloud surrounding the CND.\\

Using $^{13}$CO(3-2) observations, \citet{zylka1990} detected another streamer that is very small in spatial extent (see their Figure 5d), 
reaching from the $+$50 \KMS\textrm{ }cloud towards the CND. We detect the same feature in the \COLINEAI\textrm{ }line going from $\Delta$l $=$ $+$175'', 
$\Delta$b $=$ $-$200'' to $\Delta$l $=$ $+$225'', $\Delta$b $=$ $-$100'' between $+$45 \KMS\textrm{ }and $+$70 \KMS. The streamer is signalized in 
the right panel of Figure \ref{cii_emission_lane}, reaching towards the GC. It has a very small spatial extent and is weak in the carbon line emission. 
This streamer is also detected in the 1.3 mm dust continuum tracing free-free and thermal dust in the analysis done by \citet{zylka1998}.\\

A {western streamer} going from the H Region towards the $-$30 \KMS\textrm{ }cloud and connecting with the Arched 
Filaments (but not going through the GC) was discussed in Section \ref{molecular_clouds}. The structure runs almost parallel to the Galactic plane 
and can be seen in the \COLINEAI\textrm{ }emission between $\sim$ $-$50 \KMS\textrm{ }and $\sim$ $-$25 \KMS\textrm{ }as an almost straight line 
going from $\Delta$l $=$ $-$50'' to $\Delta$l $=$ $-$400'' at constant $\Delta$b $=$ $+$50''. Atomic carbon emission is also found towards 
some portions of the western streamer. Since this feature does not go through the GC, it could have a different origin to the streamers 
previously mentioned. Nonetheless, it could reflect gas under tidal disruption orbiting the GC.\\

\subsection{Emission around and within the CND}

\begin{figure}
\centering
\includegraphics[angle=-90, width=\hsize]{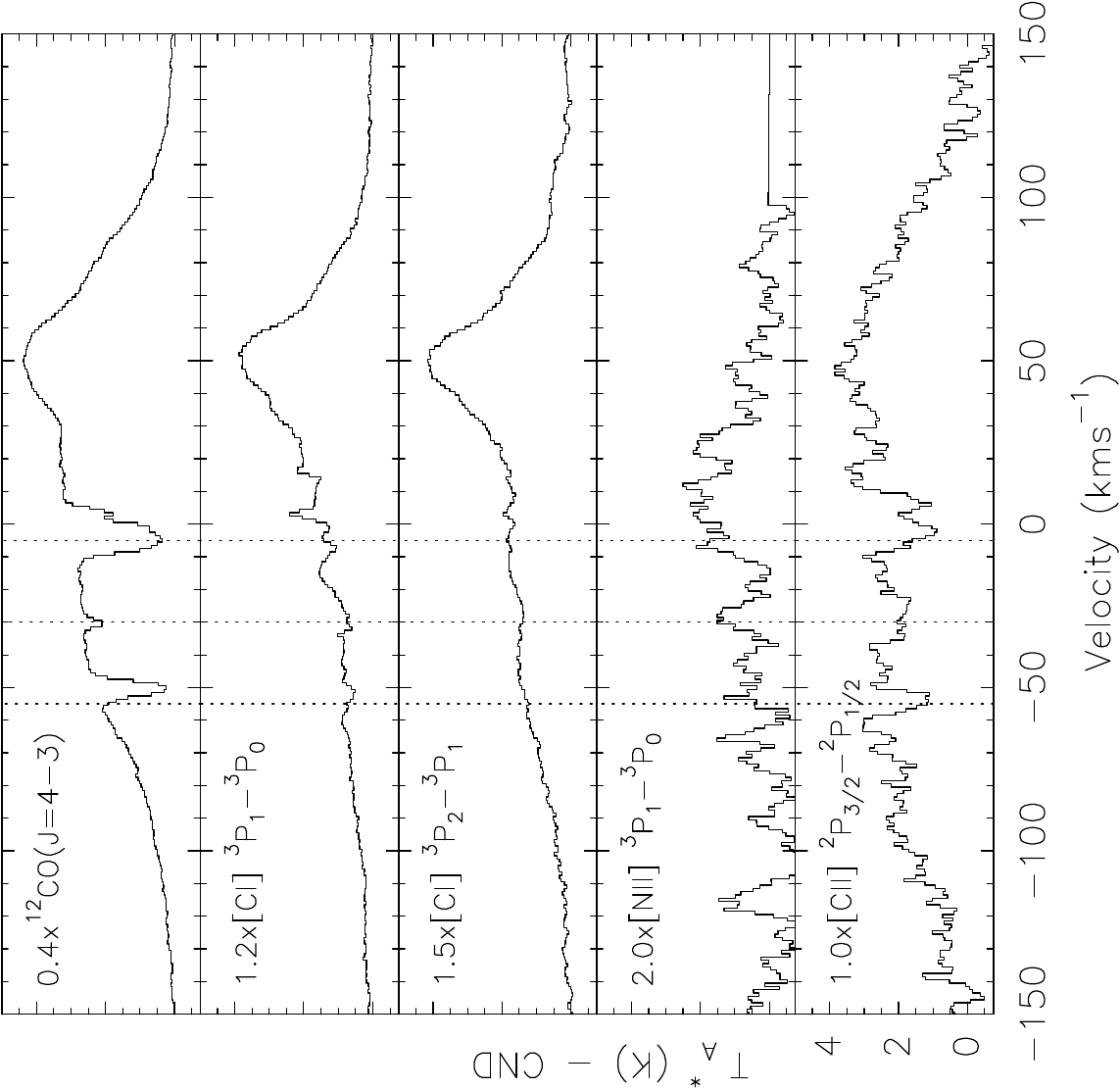}
\caption{Circumnuclear disk (CND) average spectra in the sub-mm/FIR lines presented in this work. The plots were created by averaging the spectra within a 
100''$\times$100'' area centred at the position of Sgr A$^{\star}$, from data cubes with a common 46'' spatial resolution. Vertical dotted lines show the 
loci of the 3 kpc, 4.5 kpc, and local arm spiral features at $\sim$ $-$55 \KMS, $\sim$ $-$30 \KMS, and $\sim$ $-$5 \KMS, respectively.}
\label{cnd_requena}
\end{figure}

In the Herschel-HIFI maps, the CND is contained within a square of side length 100'' centred at  $\Delta \alpha(J2000) =$ 0'', 
$\Delta \delta(J2000) =$ 0'' . This area contains the high resolution (9.5'') \mbox{CO(6-5)} emission observed by \citet{requena2012}. 
In the NANTEN/SMART \COLINEAI\textrm{ }maps measured in Galactic coordinates, this area is centred at $\Delta$l $\sim$ $-$202'', 
$\Delta$b $\sim$ $-$165''. Figure \ref{cnd_requena} shows the average emission within the 100''$\times$100'' area containing the CND. 
The emission extends from $-$150 \KMS\textrm{ }to $+$150 \KMS\textrm{ }in all lines but \NII, where the line is barely detected between 
$-$10 \KMS\textrm{ }and $+$30 \KMS. Overall, the \COLINEAI, \CILINEAI, \CILINEAII, and \CII\textrm{ }average spectra are very similar. 
All present a clear emission bump at $+$50 \KMS \ where emission from the northern lobe of the CND dominates, while towards negative velocities, 
where the southern lobe is present, the average emission decreases also showing  the strong absorption features from the loci of spiral arms 
along the l.o.s. in the molecular and ionized emission. \\

\begin{figure*}
\centering
\includegraphics[angle=-90, width=\hsize]{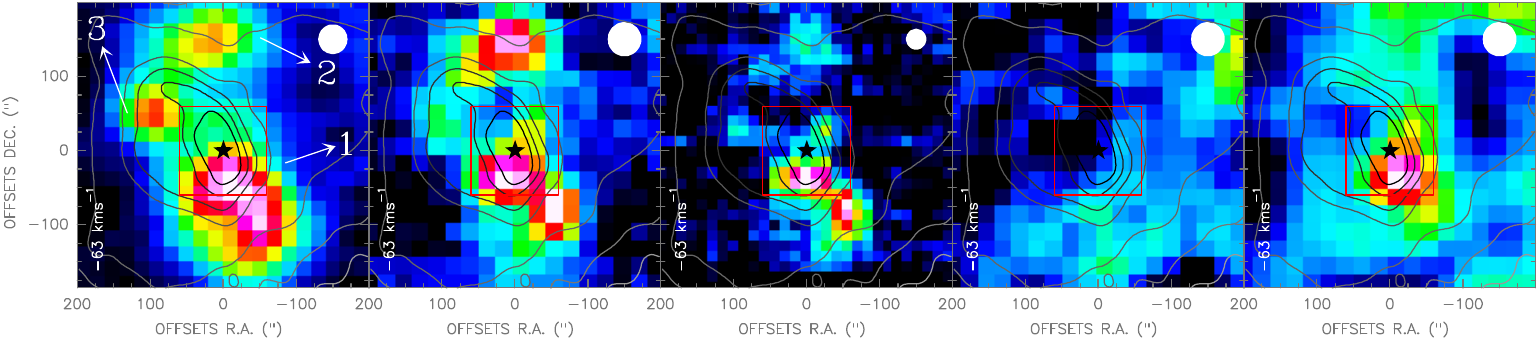}
\caption{From left to right: \COLINEAI, \CILINEAI, \CILINEAII, \NII, and \CII\textrm{ }emission at LSR velocity $-$63 \KMS\textrm{ }around and within the CND,
denoted by the red box. The 20 cm continuum observations by \citet{yusef1987a} are shown as contours. Filled white circles represent the spatial resolution 
in each panel. The $\star$ symbol represents the position of Sgr A$^{\star}$.}
\label{bright_neg_spots}
\end{figure*}

Towards negative velocities, above $-$50 \KMS, there is an excess of  \CII\textrm{ }emission with 
respect to the molecular and neutral atomic emission. This excess   traces a very bright \CII\textrm{ }spot towards 
the southern part of the CND where the peak of the integrated \mbox{CO(6-5)} intensity is also found \citep{requena2012}. This can be 
clearly seen in Figure \ref{bright_neg_spots} (number 1). The bright \CII\textrm{ }spot is 
detected in both atomic carbon lines and in the carbon monoxide line. In these transitions, two emission peaks are distinguishable, 
instead of the single emission peak in the \CII\textrm{ }emission. No significant \NII\textrm{ }emission is detected at this position. 
Outside the northern part of the CND, there is a second spot (number 2 in Figure \ref{bright_neg_spots}) at position $\Delta$l $\approx$ $-$75'', 
$\Delta$b $\approx$ $-$100'' detected in \COLINEAI\textrm{ }and in \CILINEAI\textrm{ }and \CILINEAII\textrm{ }at $\Delta \alpha(J2000) \approx$ $+$25'', 
$\Delta \delta(J20004) \approx$ $+$150''. This feature shows up for the first time at around $-$75 \KMS\textrm{ }and it is 
still visible up to $-$55 \KMS\textrm{ }where the 3 kpc absorption features appears. There is neither \NII\textrm{ }nor significant 
\CII\textrm{ }emission detected at this position. A third spot of emission (number 3 in Figure \ref{bright_neg_spots})  
located just outside the  northern part of the  CND is detected in \COLINEAI\textrm{ }at $\Delta$l $\approx$ $-$125'', $\Delta$b $\approx$ $-$220'', 
and barely detected in \CILINEAI\textrm{ }and \CILINEAII\textrm{ }at $\Delta \alpha(J2000) \approx$ $+$100'', $\Delta \delta(J20004) \approx$ $+$25'', 
traceable between $-$100 \KMS\textrm{ }and $-$55 \KMS\  before it washes out with the rest of the emission. Significant \NII\textrm{ }or 
\CII\textrm{ } emission is not detected at this position. The different atomic carbon intensities between the second and 
third spots of bright \COLINEAI\textrm{ }emission could be an indication of different excitation mechanisms for sources only 2 arcmin apart in
the vicinity of the CND. Towards very high positive LSR velocities, a similar bright spot is found associated with the northern 
lobe of the CND where the \mbox{CO(6-5)} emission also shows a local maximum \citep{requena2012}. The spot is located at $\Delta$l $\approx$ $+$25'', 
$\Delta$b $\approx$ $-$50'' in the \COLINEAI\textrm{ }line and at $\Delta \alpha(J2000) \approx$ $+$35'', $\Delta \delta(J20004) \approx$ $+$50'' 
in the carbon lines. The feature is detected in all lines, except in \NII\textrm{ }, as can be seen in the channel map at $+$105 \KMS. 

\subsection{High velocity gas detected in [CI]1-0 and CO(4-3)}

Towards the l.o.s. to the Sgr A Complex, and at high LSR velocities ($\mid$\VLSR$\mid$ $>$ 100 \KMS), the gas is expected to be orbiting in the 
family of X$_{1}$ orbits around the Galactic Centre \citep{jenkins1994}. The high velocity gas (HVG) at these orbital velocities is detected in 
low-J CO observations, while mid-J detections are difficult owing to weak emission at those frequencies. Given the low rms noise in our submillimetre 
data sets, we are able to detect this gas in the \COLINEAI\textrm{ }and the \CILINEAI\textrm{ }lines. At negative LSR velocities, HVG is detected 
between $-$160 \KMS\textrm{ }and $-$100 \KMS\textrm{ }at the lower right part of the \COLINEAI\textrm{ }channel maps and in the lower left corner 
of the \CILINEAI\textrm{ }line images. This structure seems to have two velocity components along the l.o.s, judging from
$^{13}$CO(1-0) observations of the ongoing Mopra CMZ CO(J$=$1-0) Survey (M. Burton, private communication). Between $-$96 \KMS\textrm{ }to $-$73 \KMS, a 
remarkably comet-like structure is seen in both lines. In the \COLINEAI\textrm{ }maps, it ranges from $\Delta$l $\sim$ $+$350'', $\Delta$b $\sim$ 50'' 
to $\Delta$l $\sim$ $+$650'', $\Delta$b $\sim$ $-$400''. In the \CILINEAI\textrm{ }maps, the structure spans from $\Delta \alpha(J2000) =$ $+$100'' to 
$\Delta \alpha(J2000) =$ $+$550'' at approximately constant Declination $\Delta \delta(J2000) =$ $+$600''.\\  

\begin{figure}[!h]
\centering
\includegraphics[angle=-90, width=\hsize]{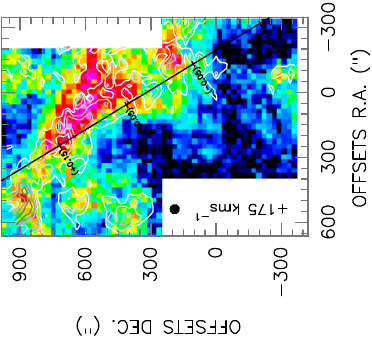}
\caption{Channel map of the \CILINEAI\textrm{ }(colour scale) and \COLINEAI\textrm{ }(contours, black is the largest value) emission of the high velocity 
gas component at $+$175 \KMS. The filled black circle shows the 46'' angular resolution of the \CILINEAI\textrm{ }data. The solid straight line shows 
the position of the Galactic plane at b $=$ 0\DEG, with marks denoting a few Galactic longitude values for orientation. The structure belongs to the X$_{1}$ 
orbits in the gravitational potential of the GC. The emission from both lines show large spatial coincidence, though intensity ratios can have large variations, 
especially towards the \COLINEAI\textrm{ }peak, at the upper left corner of the image.}
\label{hvg3_ci_co}
\end{figure}

At positive LSR velocities, we identified two large-scale structures with LSR velocities above $+$100 \KMS. The first one can be traced between LSR velocities 
$+$100 \KMS\textrm{ }to $+$150 \KMS, and is located in the upper left and upper right regions of the \COLINEAI\textrm{ }and \CILINEAI\textrm{ }channel maps, 
respectively. This emission was already detected in CS(2-1) by \citet{serabyn1987}, but no further analysis was made. At even larger LSR 
velocities, between $+$155 \KMS\textrm{ }and $+$180 \KMS, a very large structure  almost entirely covering the observed area is detected in both lines.
Figure \ref{hvg3_ci_co} shows the \CILINEAI\textrm{ }(colours) and  \COLINEAI\textrm{ }(contours) emission of this high velocity gas component 
at the $+$175 \KMS. There is a good spatial correlation between the two lines;  the carbon emission is more widespread than
the carbon monoxide. From Figure \ref{morpho:final_average_spectra}, the peak intensities in both lines are around $\sim$ 0.25 K in the average spectra. 
This structure is particularly interesting since it appears completely separated from the bulk of the gas in the X$_{2}$ orbits, 
assumed to be at radial velocities $<$ $+$100 \KMS\textrm{ }in the direction of the Sgr A Complex.\\ 

In the data sets of the neutral species, we also detect the HVCC CO$+$0.02$-$0.02 previously observed by \citet{oka2008} with high 
spatial resolution in CO(1-0) and HCN(1-0). The HVCC is clearly detected in \COLINEAI, \CILINEAI, and \CILINEAII\  between $+$100 
\KMS\textrm{ }and $+$150 \KMS. The compact source is located at the centre of the map in all observations. The \CII\textrm{ }detection 
is marginal with the emission just above the lowest contour at the  3$\sigma$ significance level, with $\sigma$ the characteristic noise of the 
data as summarized in Table \ref{tab:data_summary}.\\


\section{Summary}\label{summary}

We have presented observations of the Sgr A Complex in the sub-mm/FIR range obtained with the Herschel-HIFI satellite 
and NANTEN2/SMART telescope, tracing the warm molecular, atomic, and ionized gas within this region. The observations have high 
angular ($\leq$ 46'') and spectral resolution (1 \KMS) over a wide spectral range between $\pm$ 200 \KMS. We have discussed the 
data reduction process of the different observations and how the reference emission contamination in the Herschel-HIFI
observations was recovered from the reference position measurements after standing waves were removed, and the emission 
was modelled and added back to the observed spectra. A cross-calibration using \CILINEAI\textrm{ }emission between the NANTEN2/SMART
telescope and Herschel-HIFI satellite shows that both telescopes measure essentially the same antenna temperatures in the T$_{A}^{*}$ scale, 
which is a more appropriate scale for the measured intensities than the main beam temperature scale, given the large error beam pick 
up in both telescopes and the spatial extent of the emission in all observed lines. The main survey parameters are summarized in Table 
\ref{tab:data_summary}.\\

At negative LSR velocities, the Arched Filaments harbour most of the flux of the ionized gas in our observations. In the channel maps, the 
\CII\textrm{ }and \NII\textrm{ }emission moves from the H region, towards the Arched Filaments, first reaching  G0.07$+$0.04 where the gas 
shows signatures of interaction with the Northern Thread, and  then moving from the W1/W2 to the E1/E1 filaments and to G0.10$+$0.02. The 
\CILINEAI\textrm{ }and \CILINEAII\textrm{ }emission have local maxima at the position of the M$+$0.04$+$0.03 ($-$30 \KMS\textrm{ }cloud)
and M$+$0.02$-$0.05 ($-$15 \KMS\textrm{ }cloud), while the \COLINEAI\textrm{ }peaks  between the clouds. At positive LSR velocities 
the M2$-$0.13$-$0.08 ($+$20 \KMS\textrm{ }cloud) is traced by the carbon and carbon monoxide emission. The emission moves continuously with
increasing LSR velocities around the CND to connect with the M$-$0.02$-$0.07 ($+$50 \KMS\textrm{ }cloud). We identify this
emission lane with the eastern streamer. The $+$50 \KMS\textrm{ }cloud is  crescent-shaped in the atomic carbon lines 
(at $\sim$ $+$59 \KMS) and molecular line (at $\sim$ $+$67 \KMS) in the 3 \KMS\textrm{ }wide, integrated intensity maps. This reveals the 
interaction of the $+$50 \KMS\textrm{ }cloud with the Sgr A-East supernova remnant. In general, at positive LSR velocities, the 
distribution of the carbon emission is highly asymmetric with respect to the upper and lower parts of the Herschel-HIFI maps, 
separated at $\Delta \delta(J2000) \sim$ $+$250'';  most of bright emission is associated with the $+$50 \KMS\textrm{ }cloud. On the 
contrary, the \COLINEAI\textrm{ }emission is quite symmetric, with bright emission associated with the $+$50 \KMS\textrm{ }cloud, but of 
similar intensity at the positions tracing the molecular complex including the M$+$0.07$-$0.08, M$+$0.11$-$0.08, M$+$0.06$-$0.04, and 
M$+$0.10$-$0.01 clouds. A local maximum in the \COLINEAI\textrm{ }emission is associated with the edge of the Sickle \HII\textrm{ }region, also 
traced in the other sub-mm/FIR lines in our data. Emission around and within the CND show very extended emission in LSR velocity from $-$150 
\KMS\textrm{ }to $+$ 150 \KMS\textrm{ }in all lines but \NII. In addition, three emission spots at very large negative LSR velocities 
($\sim$ $-$63 \KMS) are found close to but outside (for the two northern spots) and within (for the southern spot) the CND, showing different 
emission strengths in our observations, which is a signature of different physical conditions. Absorption of spiral arms features along the l.o.s. 
are seen at many positions;  the most prominent is the one at $-$5 \KMS\textrm{ }associated with the loci of the local arm. 
A bright emission lane in \CII\textrm{ }is found associated with the convex side of the $+$50 \KMS\textrm{ }cloud,
with a very weak \NII\textrm{ }counterpart. Gas streamers going through the GC are clearly seen at $-$11 \KMS\textrm{ }and $+$8 
\KMS\textrm{ }connecting otherwise separated entities such as the $-$15 \KMS\textrm{ }and $+$20 \KMS\textrm{ }clouds. The similar shape 
of the streamers could be an indication that they are part of the same structure, and that the lack of emission around 0 \KMS, due to spiral 
arm absorption along the l.o.s., could artificially  put them as separated structures. A small gas streamer going from the $+$50 
\KMS\textrm{ }cloud towards the GC is also detected in our \COLINEAI\textrm{ }observations. A western streamer running parallel to the 
Galactic plane is found connecting emission associated with the H region with the one at the Arches Filaments. Although this streamer does not go
through the GC, its elongated spatial distribution could indicate gas being tidally disrupted orbiting the GC. High velocity gas 
($\mid$\VLSR$\mid$ $>$ 100 \KMS) structures associated with X$_{1}$ orbits around the GC  are detected in the \COLINEAI\textrm{ }and 
\CILINEAI\textrm{ }lines. The physical conditions of this gas are expected to differ strongly from the gas at the X$_{2}$ orbits (the bulk of the emission
in our data sets) since gas in the outer parts of the GC is subject to strong shocks on its way towards the GC. The HVCC is also seen  
in \COLINEAI, \CILINEAI, and \CILINEAII\textrm{ }emission in our observations with a marginal \CII\textrm{ }detection.\\

\begin{acknowledgements}
This work was supported by the Comisi\'on Nacional de Ciencia y Tecnolog\'ia (CONICYT), the Deutscher Akademischer Austausch Dienst (DAAD),
and the SFB956 project.\\
\end{acknowledgements}


\Online

\begin{appendix} 

\section{Baseline distortions in the \NII\textrm{ }and \CII\textrm{ }Lines}\label{appendixB}

\begin{figure}[!h]
   \centering
   \includegraphics[angle=-90, width=\hsize]{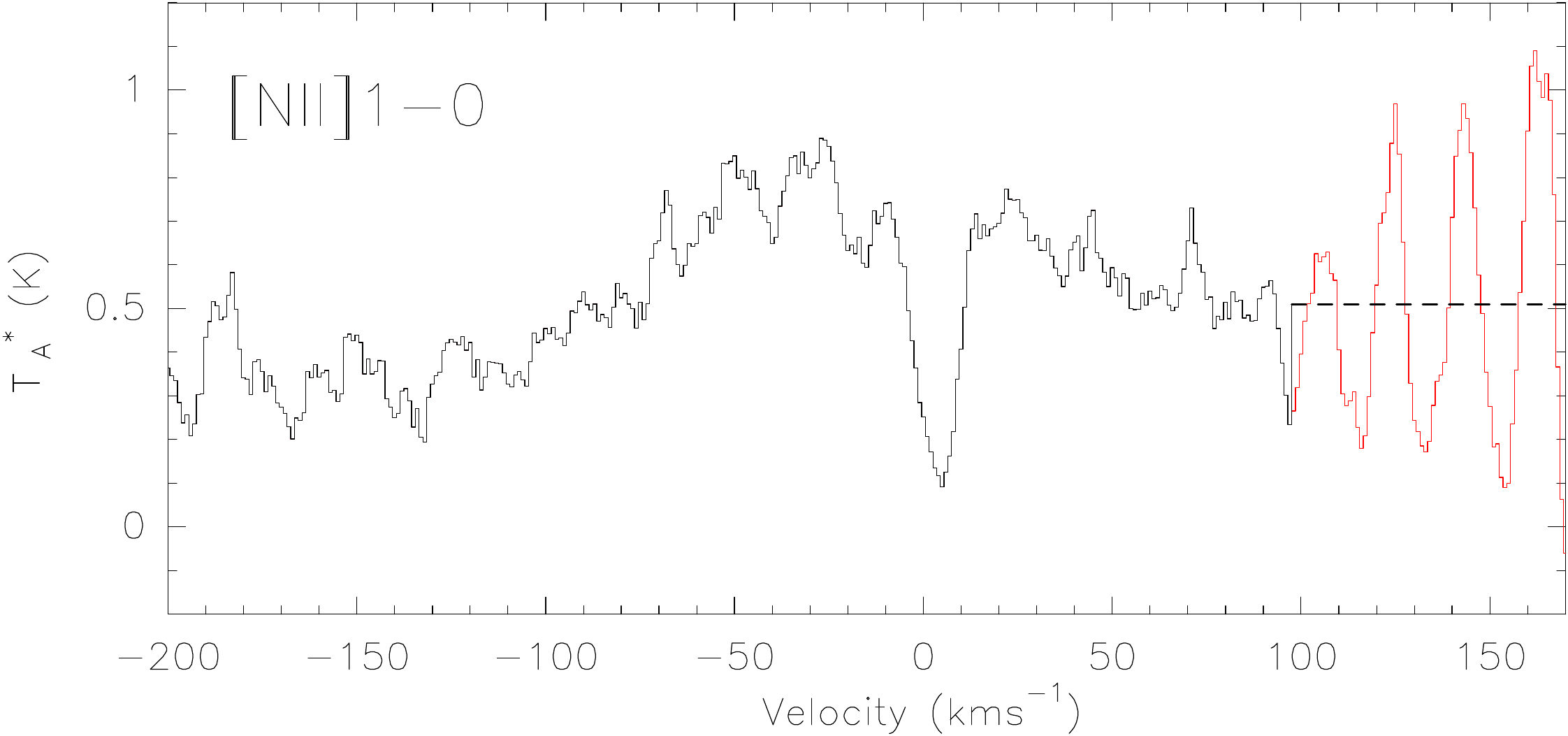}
   \caption{\NII\textrm{ }average spectrum of the V polarization observations over the whole map. The red line indicates the 
distortion of the baseline between $+$94 \KMS\textrm{ }and $+$170 \KMS\textrm{ }present in all the spectra. The dashed line over 
the same LSR velocity interval shows the result of the method utilized in getting rid of the distortion by replacing the measured 
antenna temperatures by their average value within the same LSR velocity interval.}
\label{fig_data_reduction:NII_lo_problem}
\end{figure}

\begin{figure}[!h]
   \centering
   \includegraphics[angle=-90, width=\hsize]{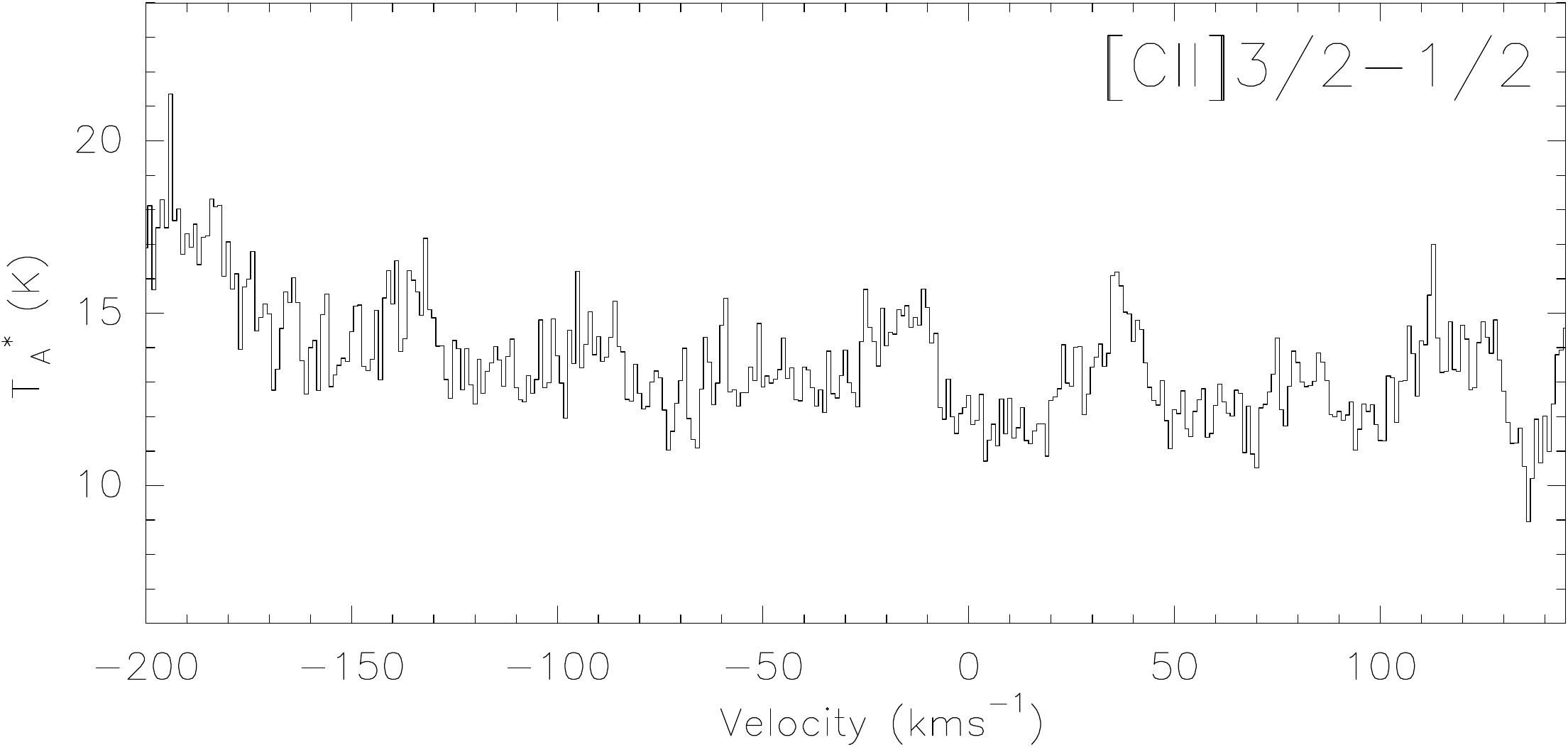}
   \caption{\CII\textrm{ }average OTF line of the V polarization spectra. The curvature of the average spectrum together with 
the distortion of the baseline across the bandpass are the product of an overpump of the V polarization mixer due to excessive 
LO power \citep{higgins2011}. Such spectra were filtered out by setting the T$_{A,rms}^{*}$ threshold to 2.3 K.}
\label{fig_data_reduction:CII_lo_problem}
\end{figure}

The baseline distortions of the \NII\textrm{ }and \CII\textrm{ }observations carried out with the HEBs mixers 
of the Herschel-HIFI satellite are shown in Figures \ref{fig_data_reduction:NII_lo_problem} and \ref{fig_data_reduction:CII_lo_problem}. 
The baseline distortion in Figure \ref{fig_data_reduction:NII_lo_problem} is produced by the combination of the diplexer coupling at this frequency, 
the electrical standing waves in the HEB mixers, and the IF frequency dependence sensitivity of the HEB mixers, which are more sensitive to
 lower IF frequencies. The red line in the figure denotes the LSR velocity range in which the average antenna temperature 
was calculated, while the dashed black line shows the result of the average. In Figure \ref{fig_data_reduction:CII_lo_problem} and average OTF
line of the \CII\textrm{ }observations illustrates the electrical standing wave patter present in the vertical polarization data
produced by an impedance mismatch between the mixer and the first low noise amplifier (LNA) due to a too high LO power pumping \citep{higgins2011}.\\

\end{appendix}
\end{document}